\documentclass{aa}
\usepackage{graphicx}
\usepackage{natbib} \bibpunct{(}{)}{;}{a}{}{,}
\usepackage{color}
\usepackage{txfonts}
\usepackage{rotating}
 
\def\changed{}
\def\changedA{}

\newcommand{\dint}{\,\mbox{d}} 

\newcommand{\kms}{\ifmmode{\,\mbox{km}\,\mbox{s}^{-1}}\else{km/s}\fi}
\newcommand{\msun}{\ifmmode M_{\odot} \else M$_{\odot}$\fi}
\newcommand{\rsun}{\ifmmode R_{\odot} \else R$_{\odot}$\fi}
\newcommand{\lsun}{\ifmmode L_{\odot} \else L$_{\odot}$\fi}
\newcommand{\zsun}{\ifmmode Z_{\odot} \else $Z_{\odot}$\fi}
\newcommand{\velo}{\ifmmode\varv\else$\varv$\fi}
\newcommand{\vinf}{\ifmmode\velo_\infty\else$\velo_\infty$\fi}
\begin{document} 

\title{Stellar envelope inflation near the Eddington limit} 
\subtitle{Implications for the radii of Wolf-Rayet stars and luminous blue variables}
 
\titlerunning{Stellar envelope inflation near the Eddington limit}
 
\author{G.\ Gr\"{a}fener\inst{\ref{inst1}}
\and    S.P.\ Owocki\inst{\ref{inst2}}
\and    J.S.\ Vink\inst{\ref{inst1}}
}

\institute{Armagh Observatory, College Hill, Armagh, BT61\,9DG, United
  Kingdom\label{inst1} \and Bartol Research Institute, University of
  Delaware, Newark, DE 19716, USA\label{inst2} }

\date{Received ; Accepted}
 
\abstract{It has been proposed that the envelopes of luminous stars
  {{\changedA may}} be subject to substantial radius inflation.
  The peculiar structure of such inflated envelopes, with an almost
  void, radiatively dominated region beneath a thin, dense shell could
  mean that many in reality compact stars are hidden below inflated
  envelopes, displaying much lower effective temperatures.  {{\changedA
      The inflation effect has been discussed in relation to the
      radius problem of Wolf-Rayet (WR) stars, but has yet failed to
      explain the large observed radii of Galactic WR stars.}}  } { We
  wish to obtain a physical {{\changedA perspective of the inflation
      effect}}, and study the consequences for the radii of Wolf-Rayet
  (WR) stars, and luminous blue variables (LBVs).  For WR stars the
  observed radii are up to an order of magnitude larger than predicted
  by theory, whilst S Doradus-type LBVs are subject to humongous
  radius variations, which remain as yet ill-explained.}  {{{\changedA
      We use a dual approach to investigate the envelope inflation,
      based on numerical models for stars near the Eddington limit,
      and a new analytic formalism to describe the effect.  An
      additional new aspect is that we take the effect of density
      inhomogeneities (clumping) within the outer stellar envelopes
      into account.}}}
{{{\changedA Due to the effect of clumping we are able to bring the
      observed WR radii in agreement with theory.  Based on our new
      formalism, we find that the radial inflation is a function of a
      dimensionless parameter $W$, which largely depends on the
      topology of the Fe-opacity peak, i.e., on material properties.}}
  For $W>1$, we {{\changedA discover}} an instability limit, for which
  the stellar envelope becomes gravitationally unbound, i.e.\ there no
  longer exists a static solution.  Within this framework we
  {{\changedA are also able to explain the S\,Doradus-type
      instabilities}} for LBVs like AG\,Car, with a possible
  triggering due to changes in stellar rotation.  }
{The stellar effective temperatures in the upper Hertzsprung-Russell
  (HR) diagram are potentially strongly affected by the inflation
  effect.  This may have particularly strong effects on the evolved
  massive LBV and WR stars just prior to their final collapse, as the
  progenitors of supernovae (SNe)\,Ibc, SNe\,II, and long-duration
  gamma-ray bursts (long GRBs).}

\maketitle

\section{Introduction} 
\label{sec:intro} 

In the standard picture for the evolution of the most massive stars
(with $M \ga 30\,\msun$), O stars have been proposed to evolve through
successive luminous blue variable (LBV) and Wolf-Rayet (WR) phases
before exploding as hydrogen (H) free supernovae (SNe) Ibc
\citep[e.g.,][]{lan1:94,mey1:03,heg1:03}.  This evolutionary path is
however still open as LBVs have more recently been suggested to be the
direct progenitors of some H-rich type II SNe
\citep{kot1:06,smi2:07,gal1:09}. Additional renewed interest in WR
stars arises from their connection to long-duration gamma ray bursts
\citep[long GRBs, cf.][]{yoo1:05,woo1:06}.

What LBVs and WR stars have in common is their proximity to the
Eddington limit.  Not only is this thought to be instrumental for
their mass-loss behavior \citep{vin1:02,gra1:08,vin1:11,gra1:11}, but
it may also have key consequences for their stellar structures.  Using
modern OPAL opacities, \citet{ish1:99}, \citet{pet1:06}, and
\citet{yun1:08} studied the interior structure of massive stars,
revealing a ``core-halo'' configuration that involves a relatively small
convective core and an extended radiative envelope. This is referred
to as the ``inflation'' of the outer envelope.

It has been known for many decades that the observed radii of WR stars
are almost an order of magnitude larger than canonical WR stellar
structure models indicate. This is oftentimes attributed to their
so-called pseudo-photospheres, which concern an ``effective''
photosphere several times larger than the hydrostatic radius, as a
result of a dense stellar outflow \citep[e.g.,][]{cro1:07}.
Pseudo-photospheres have also been discussed in the context of LBVs,
although the issue is still under debate \citep[e.g.,][]{smi1:04}.

An alternative explanation for the large radii of WR stars may involve
envelope inflation. The issue is particularly relevant in the context
of WR stars as GRB progenitors, as there have been several suggestions
that the required radii of GRB progenitors are up to an order of
magnitude ($\sim$\,$1\,R_\odot$ vs. $\sim$\,$10\,R_\odot$) smaller
than the radii of observed WR stars \citep{mod1:09,cui1:10}.  The WR
radius issue is thus highly relevant for GRB modeling.

For LBVs the issue of their radii is equally interesting. The defining
property of LBVs concerns their ``S Doradus'' cycles. On timescales of
years to decades LBVs are seen to vary between effective temperatures
of $\sim$30\,kK (early B spectral type) to $\sim$8\,kK (early F). Due
to the lack of convincing counter-evidence, these S Dor excursions in
the upper Hertzsprung-Russell diagram have generally been assumed to
occur at constant bolometric luminosity \citep{hum1:94}, but recent
studies have challenged this behavior for a couple of objects
\citep{gro1:09,cla1:09}. Either way, the issue of the increased LBV
radii during S Dor cycles remains un-questioned, but a satisfying
explanation for it has yet to be provided \citep{vin1:09}.  Outer
envelope inflation may turn out to be an interesting explanation for S
Dor variations, as will be detailed in the following.

The paper is organized as follows.  In Sect.\,\ref{sec:modelcomp} we
present numerical stellar structure models that show the envelope
inflation effect. In Sect.\,\ref{sec:anal} we describe the effect
analytically, {\changedA including a new instability limit. We also
  provide a recipe that can be used to estimate the inflation effect
  for arbitrary stars with given (observed) parameters}. The
implications of our results are discussed in
Sect.\,\ref{sec:implications}, with focus on WR\,stars and LBVs. The
conclusions are summarized in Sect.\,\ref{sec:conclusions}.

\section{Model computations}
\label{sec:modelcomp}

In this section we present stellar structure models for chemically
homogeneous stars close to the Eddington limit. To model the
conditions in classical WN-stars (i.e., WR stars in the phase of core
He-burning that have lost their H-rich envelope), we use pure
He-models, while models with hydrogen resemble the conditions in
massive WNh stars, and LBVs. At this point we note that we only use
the assumption of chemical homogeneity, to keep the models as simple
as possible. The occurrence of an envelope inflation is not connected
to this assumption, as it only affects the outermost layers of the
star.

In Sect.\,\ref{sec:numerics} we give an overview of our numerical
approach for computing the stellar structure, and in
Sect.\,\ref{sec:inflation} we describe the envelope inflation, as it
occurs in our models. In Sect.\,\ref{sec:grid}, we present a model
grid to investigate the occurrence of this effect in the HR diagram.
Finally, in Sect.\,\ref{sec:clumping} we discuss the important
influence of density inhomogeneities on our models.

\subsection{Numerical method}
\label{sec:numerics} 

Our models are computed with a code that integrates the stellar
structure equations using a shooting method.  The numerical methods
are described in the textbook by \citet{han1:94}.  The code is based
on an example program that is distributed with the book, but is
completely re-written, and updated with OPAL opacities
\citep{igl1:96}. {\changedA As the chemical structure of the stars is
  fixed, a set of four differential equations needs to be solved.}
These are the equations of hydrostatic equilibrium
\begin{equation}
\frac{dP}{dm} = -\frac{Gm}{4\pi r^4},
 \label{eq:hydeq}
\end{equation}
mass conservation
\begin{equation}
\frac{dr}{dm} = \frac{1}{4\pi r^2 \rho},
 \label{eq:mass}
\end{equation}
energy conservation
\begin{equation}
\frac{dL}{dm} = \epsilon,
 \label{eq:econs}
\end{equation}
and energy transport
\begin{equation}
\frac{d\ln T}{d\ln P} = \nabla,
 \label{eq:etrans}
\end{equation}
where $\nabla = \nabla_{\rm rad}$ for the case of purely radiative
energy transport, and $\nabla_{\rm rad} > \nabla > \nabla_{\rm ad}$
for the convective case. {\changed As our models are pure stellar
  structure models, i.e.\ they include no time-dependent terms, the
  energy generation rate $\epsilon$ in Eq.\,\ref{eq:econs} does not
  take changes in the gravitational energy, due to
  contraction/expansion of the star into account. This omission has no
  consequences for the physics of the low-density envelopes discussed
  in this work, but can potentially affect the L/M ratio of WR stars
  in late evolutionary stages (see Sect.\,\ref{sec:DYN} for a detailed
  discussion).}

{\changed Furthermore, we neglect dynamic terms due to mass loss,
  i.e.,} we only concentrate on the hydrostatic case. The dynamic case
has been discussed by \citet{pet1:06}. For the case of WR stars with
strong winds, \citeauthor{pet1:06} found that dynamic effects may
inhibit the formation of inflated envelopes. We will address this
point in more detail in {\changed Sect.\,\ref{sec:MDOT}.}

The choice of the outer boundary condition, particularly the
temperature at the outer boundary, turns out to be important for the
formation of inflated envelopes. {\changed Here,} we prescribe $\rho$,
and $T$ at the outer boundary of the star in a way, that resembles
reasonable values for the sonic points of stars with strong,
radiatively driven winds.

An envelope inflation only occurs in our models if we choose outer
boundary temperatures below $\sim$\,70\,kK, well below the temperature
of the hot Fe-opacity peak.  Such temperatures are expected at the
sonic points of optically thick, radiatively driven winds, such as the
late-type WN stars described by \citet{gra1:08}.  The density $\rho$
is chosen such that the resulting mass loss rate,
\begin{equation}
  \dot{M} = 4\pi\,\rho\,a\,R_\star^2,
\end{equation}
lies in a realistic range. 

{\changed We note that our envelope solutions are very robust with
  respect to the detailed choice of $\rho$, and $T$, as long as the
  temperature $T$ lies clearly below the regime of the hot Fe-opacity
  peak, which has its maximum at $\sim 160$\,kK. The effect of changes
  in the outer boundary condition is discussed in more detail in
  Sect.\,\ref{sec:OB}.}

The stellar temperatures $T_\star$ given in this work, are effective
temperatures related to the outer boundary radius $R_\star$ of our
models, i.e., $T_\star = (L/4\pi \sigma R_\star^2)^{1/4}$. For static
atmospheres these values will be almost identical to classical
effective temperatures $T_{\rm eff}$. In the presence of dense,
extended stellar winds, our $T_\star$ values will resemble the
``effective core temperatures'' $T_\star$, that are typically inferred
from models for expanding atmospheres \citep[cf.][]{gra1:02}.

\begin{figure}[t!]
  \parbox[b]{0.49\textwidth}{\center\includegraphics[scale=0.45]{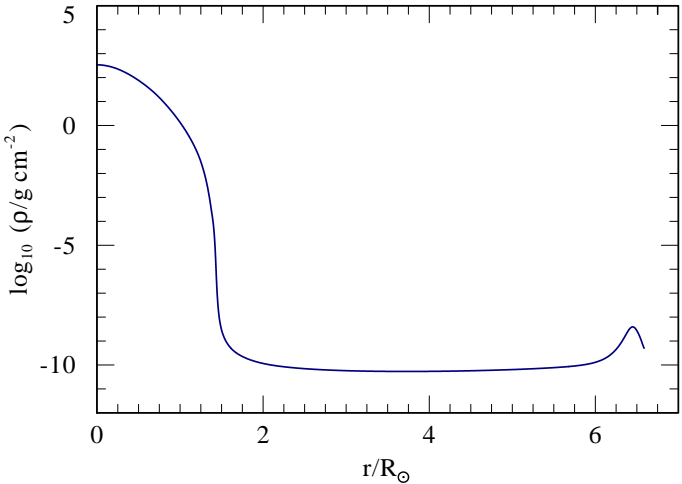}}
  \caption{Density vs.\ radius for our $23\,M_\odot$ He model
    (cf.\,Table\,\ref{tab:modparHE}).}
  \label{fig:density}
\end{figure}

\begin{figure}[t!]
  \parbox[b]{0.49\textwidth}{\center\includegraphics[scale=0.45]{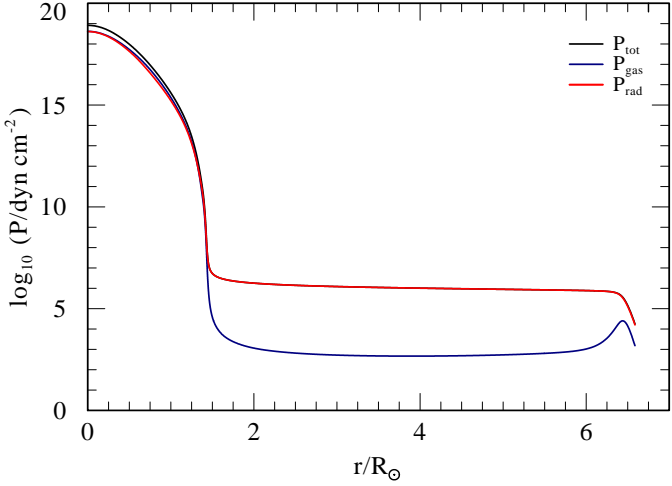}}
  \caption{Pressure vs.\ radius. Gas pressure $P_{\rm gas}$, radiation
    pressure $P_{\rm rad}$, and total pressure $P_{\rm tot}=P_{\rm
      gas}+P_{\rm rad}$.}
  \label{fig:pressure}
\end{figure}

\begin{figure}[t!]
  \parbox[b]{0.49\textwidth}{\center\includegraphics[scale=0.45]{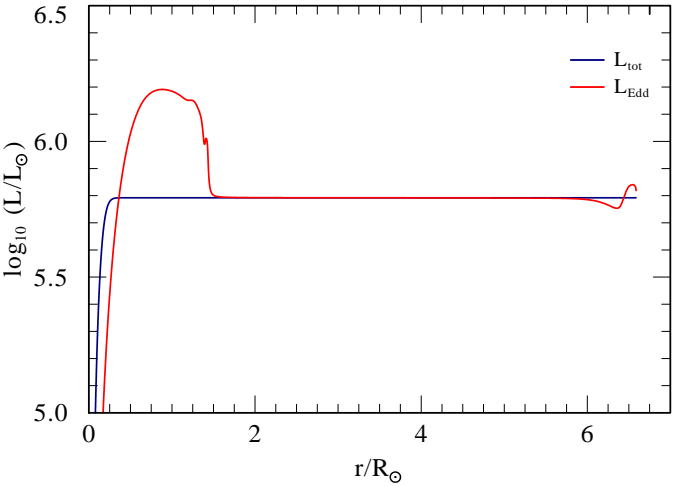}}
  \caption{Total stellar luminosity $L(r)$, compared to the Eddington
    luminosity $L_{\rm Edd} = 4\pi c G M / \kappa$.}
  \label{fig:lumi}
\end{figure}

\begin{figure}[t!]
  \parbox[b]{0.49\textwidth}{\center\includegraphics[scale=0.45]{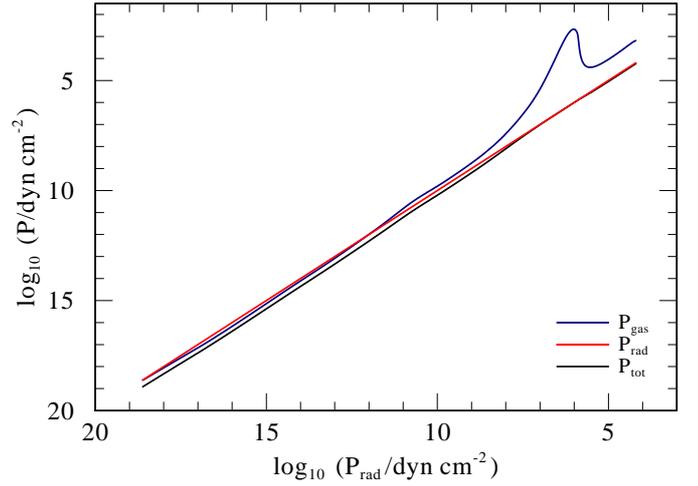}}
  \caption{Stellar structure in the $P_{\rm rad}$--$P_{\rm gas}$
    plane. Gas pressure $P_{\rm gas}$, radiation pressure $P_{\rm
      rad}$, and total pressure $P_{\rm tot}=P_{\rm gas}+P_{\rm rad}$
    are plotted vs. $P_{\rm rad}$ throughout the whole star.}
  \label{fig:pgasprad}
\end{figure}

\begin{figure*}[t!]
  \parbox[b]{0.99\textwidth}{\center{\includegraphics[scale=0.5]{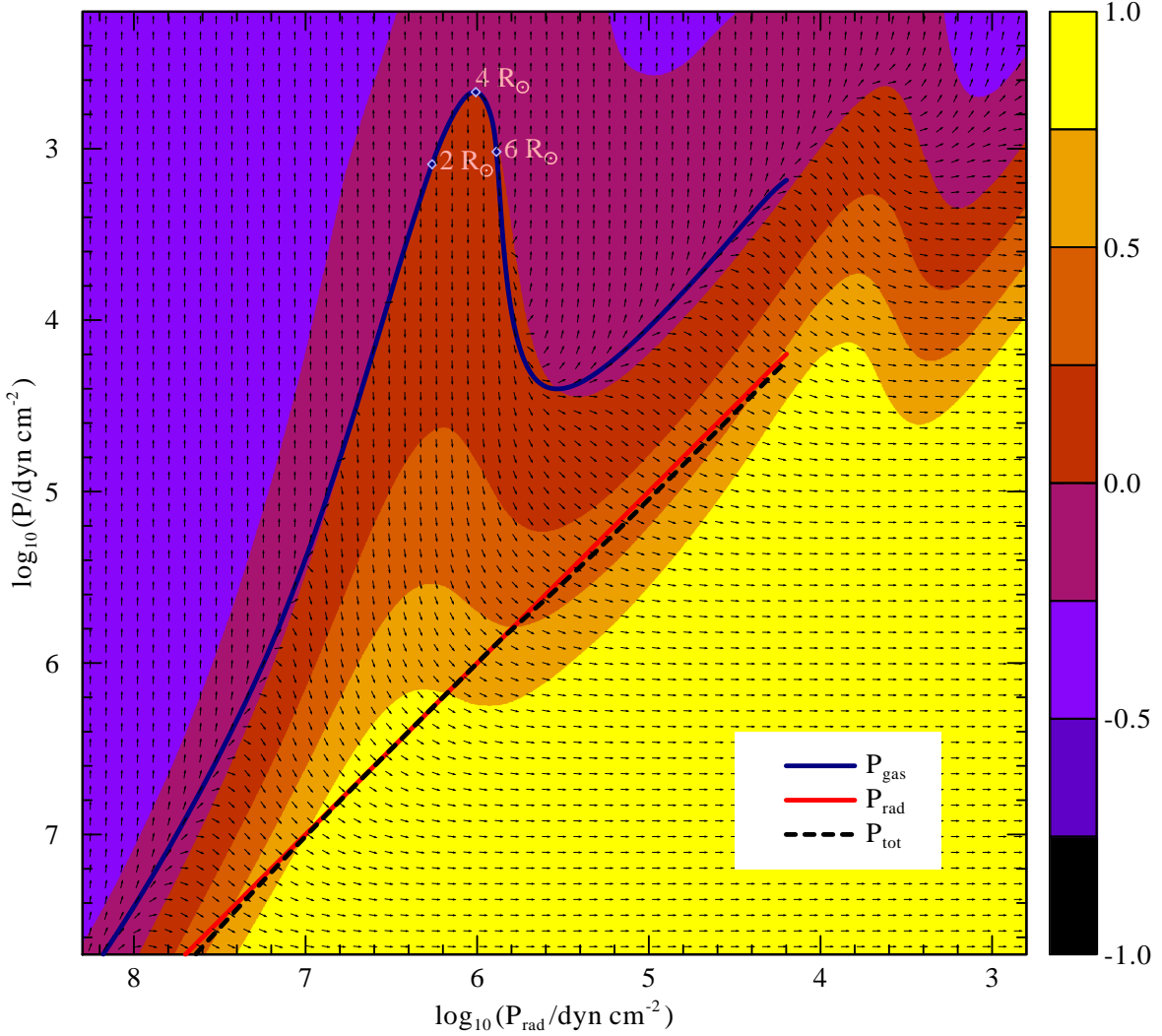}}}
  \caption{Envelope solution in the $P_{\rm rad}$--$P_{\rm gas}$
    plane. The colors indicate the logarithm of the Eddington factor
    $\Gamma$, for given $P_{\rm rad}$ and $P_{\rm gas}$, according to
    the OPAL opacity tables \citep{igl1:96}, for our pure He model
    with $23\,M_\odot$. The envelope solution almost precisely follows
    a path with $\Gamma = 1$. The radii of 2, 4, and $6\,R_\odot$
    within the inflated envelope are indicated. Arrows indicate the
    slope of the solution, as expected from Eq.\,\ref{eq:stan}.}
  \label{fig:gamma}
\end{figure*}

\subsection{Stellar structure models with inflated envelopes}
\label{sec:inflation}

As a typical example for a model with an inflated envelope, we present
a homogeneous stellar structure model for a 23\,$M_\odot$ helium star
with solar metallicity ($X=0$, $Y=0.98$, $Z=0.02$). In agreement with
previous works by \citet{ish1:99}, and \citet{pet1:06}, this model
develops an extended low-density envelope, with a density inversion
close to the surface (cf.\ Fig.\,\ref{fig:density}). The star thus
forms a shell around the actual stellar core, with a radius that is
3--4 times larger than the core radius.

Fig.\,\ref{fig:pressure} shows that the inflated region is dominated
by radiation pressure. In this model, the gas pressure $P_{\rm gas}$
only accounts for a fraction of $\sim$\,$4$$\cdot$$10^{-4}$ of the total
pressure $P$. In contrast to this, $P_{\rm gas}$ and $P_{\rm rad}$ are
roughly equal in the stellar core. The core is thus only fairly close
to the Eddington limit, as reflected by an Eddington parameter of
$\Gamma_{\rm e}\approx 0.4$ (cf.\ Table\,\ref{tab:modparHE}).  Here
$\Gamma_{\rm e}$ denotes the ``classical'' Eddington parameter, which
is related to the Thomson opacity $\kappa_{\rm e}$ due to free
electrons, i.e., $\Gamma_{\rm e} = \kappa_{\rm e} L_{\rm rad}/(4\pi c
G M)$. {\changed The $\Gamma_{\rm e}$ given in
  Tables\,\ref{tab:modparHE} and \ref{tab:modparH}, are computed for a
  fully ionized plasma with a given hydrogen mass fraction $X$, and
  $L_{\rm rad}=L$. $\Gamma_{\rm e}$ then simplifies to}
\begin{equation}
  \label{eq:gammae}
  \log(\Gamma_{\rm e}) = -4.813 + \log(1+X)
  + \log(L/L_\odot)
  - \log(M/M_\odot).
\end{equation}

In contrast, the {\changed total Eddington factor $\Gamma = \kappa
  L_{\rm rad}/(4\pi c G M) $, which is} related to the total opacity
$\kappa$, approaches unity in the inflated zone. This is shown in
Fig.\,\ref{fig:lumi}, where we compare the total luminosity within our
model with the Eddington luminosity $L_{\rm Edd} = 4\pi c G M /
\kappa$ (the condition $L_{\rm rad}=L_{\rm Edd}$ is thus equivalent to
$\Gamma = 1$). In Fig.\,\ref{fig:lumi} we can identify three zones
within the star. In the {\em inner, convective core,} the total
luminosity $L$ is larger than the Eddington luminosity $L_{\rm Edd}$.
The main reason for this is that the energy transport is almost
entirely convective, and thus $L_{\rm rad}$ is very low.  Above this
region follows the {\em radiative envelope} with $L < L_{\rm Edd}$.
On top of this follows the {\em inflated zone,} where $L_{\rm Edd}$
almost equals $L$. $\Gamma$ is thus extremely close to one, almost
throughout the complete outer envelope of the star.  This is
surprising, because $\Gamma$ is a function of the opacity $\kappa$,
which is a strongly varying function of $\rho$, and $T$. The
conditions in the inflated envelope thus demand for a mechanism that
adjusts $\Gamma$ very precisely to one.

{\changed To achieve a large radial extension it is necessary that the
  density scale height $H$ is comparable to the stellar core radius
  $R_{\rm c}$. For our example model we have $R_{\rm c}/H =
  GM(1-\Gamma)/(a^2 R_{\rm c}) \approx 3$$\cdot$$10^{3}(1-\Gamma)$. From
  Fig.\,\ref{fig:pressure} we can infer that $(1-\Gamma) \approx
  P_{gas}/P \approx 4$$\cdot$$10^{-4}$ (cf.\ Eq.\,\ref{eq:PGAM}).  Due to
  the low ratio of gas pressure to radiation pressure, $\Gamma$ is
  thus just close enough to one, to achieve a large radial extension.}

A qualitative idea about the reason for the envelope inflation can be
obtained from Fig.\,\ref{fig:pgasprad}, where we plot the gas pressure
$P_{\rm gas}$, and the total pressure $P$, vs.\ the radiation pressure
$P_{\rm rad}$.  This diagram is particularly useful, as for stars
close to the Eddington limit, $P$ is of the same order of magnitude as
$P_{\rm rad}$, i.e., in a logarithmic diagram the stellar structure
just follows a straight line (in accordance with the Eddington
standard model \citep{edd1:18} for which $P_{\rm rad}/P = const.$ is
adopted). For $P_{\rm gas}$ the situation is more complicated.  In the
stellar core, $P_{\rm gas}$ follows a similar relation as $P$.  In the
extended envelope, $P_{\rm gas}$ however drops significantly, and
rises again after a certain minimum pressure $P_{\rm min}$ is reached.
As demonstrated in Fig.\,\ref{fig:gamma}, the region where $P_{\rm
  gas}$ drops corresponds precisely to the location of the Fe-opacity
peak in the $P_{\rm rad}$--$P_{\rm gas}$ plane.

More explicitly, starting from the stellar core, the solution follows
a straight line with $P_{\rm gas}/P_{\rm rad} = const.$ This naturally
leads into a region in the $P_{\rm rad}$--$P_{\rm gas}$ plane, where
the opacity $\kappa$ increases significantly, due to the presence of
the Fe-opacity peak. When $\kappa$ reaches $\kappa_{\rm Edd}$, this
leads to $\Gamma \rightarrow 1$.  To avoid a super-Eddington
situation, with $\Gamma > 1$, either $L_{\rm rad}$, or $\kappa$ need
to be reduced.  $L_{\rm rad}$ could in principle be reduced by
convection, however, in our case the involved densities turn out to be
too low, i.e., convection is inefficient
(cf.\,Sect.\,\ref{sec:conveff}).  The only possibility to reduce
$\kappa$, is thus to lower the density.  For this reason the density
has to drop in our model, until the tip of the Fe-peak is reached (cf.
Fig.\,\ref{fig:gamma}). At the same time, however, this leads to a
situation where $P_{\rm gas} \ll P_{\rm rad}$, {\changed and thus to
  $\Gamma \rightarrow 1$ (because, as already mentioned above
  $(1-\Gamma) \approx P_{\rm gas}/P$).}  This mechanism thus forces
the solution to follow a path with $\Gamma = 1$, i.e., with
$\kappa(\rho, T) = \kappa_{\rm Edd}$.

As $P_{\rm rad}$, and $P_{\rm gas}$ are functions of $\rho$, and $T$,
we can express $\kappa$ in the form $\kappa(P_{\rm gas}, P_{\rm
  rad})$. This is actually done in Fig.\,\ref{fig:gamma}, where the
colors indicate $\kappa(P_{\rm gas}, P_{\rm rad})/\kappa_{\rm Edd}$ on
a logarithmic scale, as obtained from the OPAL opacity tables
\citep{igl1:96}.  In this plot we can see that the solution within the
low-density zone is completely dominated by the topology of the
opacity $\kappa$, i.e., by material properties.  As soon as the
solution reaches low densities close to the Fe-opacity peak, it just
follows a path with $\Gamma =1$, until a minimum density $\rho_{\rm
  min}$ is reached.  This minimum is located at the tip of the
Fe-opacity peak, typically at $P_{\rm rad} \approx 10^6\,{\rm
  dyn\,cm}^{-2}$, corresponding to a temperature of $\sim$\,150\,kK.
After the opacity peak is passed, the solution still follows $\Gamma
=1$, but now with increasing $P_{\rm gas}$, leading to a density
inversion.

To maintain the described envelope inflation, it is thus necessary
that 1) the star is close enough to the Eddington limit, so that
$\Gamma \rightarrow 1$, and 2) convection is inefficient.  If the
envelope inflation occurs, its properties are determined by the
topology of the Fe-opacity peak in the $P_{\rm rad}$--$P_{\rm gas}$
plane, i.e., by material properties. In Sect.\,\ref{sec:anal}, this
will help us to find an analytical description of the inflation
effect. It also implies a dependence on material properties like the
metallicity $Z$, as discussed by \citet{ish1:99}, and \citet{pet1:06}.
In Sect.\,\ref{sec:clumping}, we will discuss an additional important
effect, namely the influence of density inhomogeneities within the
inflated zone, which is based on a similar principle.

\subsection{A grid of homogeneous stellar structure models with
  envelope inflation}
\label{sec:grid}

\begin{table*}
  \caption{Model parameters for pure helium models. \label{tab:modparHE}}
  \centering 
  {\tiny
  \begin{tabular}{rllllllllllllllll} 
    \hline \hline 
    \rule{0cm}{2.8ex} $M_\star$ & $L_\star$ & $\Gamma_{\rm e}$ & $\log(T_{\rm c})$ & $\log(T_\star)$ & $R_{\rm c}$ & $R_{\rm e}$ & $R_\star$ & $\Delta R$ & $\Delta R$ & $\Delta M$ & $\rho_{\rm min}$ & $f_\rho$ & $Q$ & $W_c$ & $W_e$ & $P_{\rm e}$ \\
  $[M_\odot]$ & $[L_\odot]$ & & $[{\rm K}]$ & $[{\rm K}]$ & $[R_\odot]$ & [$R_\odot$] & $[R_\odot]$ & $[R_{\rm c}]$ & $[R_{\rm e}]$ & $[M_\odot]$ & $[\frac{\rm g}{{\rm cm}^3}]$ & & & & & $[\frac{\rm dyn}{{\rm cm}^2}]$ \\
\hline
\multicolumn{8}{l}{\rule{0cm}{2.2ex} $X=0.00$, $D=1$}\\
\hline
\rule{0cm}{2.2ex}
   23.7&  5.813 &  0.414 &   5.11 &   4.61 &   1.6 &   14.6 &   16.5 &   7.99 &   0.89 & $1.9$$\cdot$$10^{-6}$ & $4.4$$\cdot$$10^{-11}$ &   2.4 &  30.50 &   0.89 &   7.99 & $6.8$$\cdot$$10^5$ \\
   23.5&  5.807 &  0.412 &   5.10 &   4.68 &   1.7 &   10.5 &   11.4 &   5.35 &   0.84 & $4.6$$\cdot$$10^{-7}$ & $4.6$$\cdot$$10^{-11}$ &   2.1 &  25.27 &   0.84 &   5.35 & $6.8$$\cdot$$10^5$ \\
   23.0&  5.792 &  0.406 &   5.10 &   4.80 &   1.6 &    6.3 &    6.6 &   2.88 &   0.74 & $5.7$$\cdot$$10^{-8}$ & $5.4$$\cdot$$10^{-11}$ &   2.1 &  21.90 &   0.74 &   2.88 & $6.9$$\cdot$$10^5$ \\
   22.0&  5.761 &  0.396 &   5.11 &   4.92 &   1.6 &    3.6 &    3.7 &   1.35 &   0.57 & $6.6$$\cdot$$10^{-9}$ & $7.3$$\cdot$$10^{-11}$ &   2.0 &  16.61 &   0.57 &   1.35 & $7.1$$\cdot$$10^5$ \\
   21.0&  5.729 &  0.385 &   5.11 &   4.98 &   1.5 &    2.7 &    2.7 &   0.79 &   0.44 & $2.0$$\cdot$$10^{-9}$ & $9.9$$\cdot$$10^{-11}$ &   2.0 &  12.69 &   0.44 &   0.79 & $7.1$$\cdot$$10^5$ \\
   20.0&  5.694 &  0.373 &   5.11 &   5.01 &   1.4 &    2.2 &    2.2 &   0.52 &   0.34 & $9.6$$\cdot$$10^{-10}$ & $1.4$$\cdot$$10^{-10}$ &    2.1 &  10.28 &   0.34 &   0.52 & $7.4$$\cdot$$10^5$ \\
   19.0&  5.658 &  0.361 &   5.11 &   5.04 &   1.4 &    1.9 &    1.9 &   0.36 &   0.26 & $5.3$$\cdot$$10^{-10}$ & $1.9$$\cdot$$10^{-10}$ &    2.1 &   7.80 &   0.26 &   0.36 & $6.9$$\cdot$$10^5$ \\
   18.0&  5.618 &  0.348 &   5.11 &   5.06 &   1.3 &    1.6 &    1.7 &   0.25 &   0.20 & $3.6$$\cdot$$10^{-10}$ & $2.7$$\cdot$$10^{-10}$ &    2.1 &   5.90 &   0.20 &   0.25 & $6.9$$\cdot$$10^5$ \\
   17.0&  5.576 &  0.335 &   5.11 &   5.07 &   1.3 &    1.5 &    1.5 &   0.19 &   0.16 & $2.7$$\cdot$$10^{-10}$ & $3.9$$\cdot$$10^{-10}$ &    2.3 &   4.81 &   0.16 &   0.19 & $7.0$$\cdot$$10^5$ \\
   16.0&  5.531 &  0.320 &   5.10 &   5.07 &   1.2 &    1.4 &    1.4 &   0.14 &   0.12 & $2.1$$\cdot$$10^{-10}$ & $5.5$$\cdot$$10^{-10}$ &    2.3 &   3.76 &   0.12 &   0.14 & $6.8$$\cdot$$10^5$ \\
   15.0&  5.482 &  0.305 &   5.10 &   5.08 &   1.2 &   -- &   1.3 &   --  &   --  &   --  &   --  &   -- &   3.10 &   --  &  --   &   --  \\
   14.0&  5.429 &  0.289 &   5.10 &   5.08 &   1.1 &   -- &   1.2 &   --  &   --  &   --  &   --  &   -- &   2.31 &   --  &  --   &   --  \\
   12.0&  5.307 &  0.255 &   5.09 &   5.07 &   1.0 &   -- &   1.1 &   --  &   --  &   --  &   -- &   -- &   1.27 &   --  &  --   &   --  \\
   10.0&  5.155 &  0.216 &   5.07 &   5.06 &   0.9 &   -- &   0.9 &   --  &   --  &   --  &   -- &   -- &   0.69 &   --  &  --   &   --  \\
   8.0&  4.959 &  0.172 &   5.05 &   5.04 &   0.8 &   -- &   0.8 &   --  &   --  &   --  &   -- &   -- &   0.36 &   --  &  --   &   --  \\
\hline
\multicolumn{8}{l}{\rule{0cm}{2.2ex} $X=0.00$, $D=4$}\\
\hline
\rule{0cm}{2.2ex}
  18.5 &  5.638 &  0.355 &   5.10 &   4.69 &   1.4 &   8.1 &   9.2 &   4.82 &   0.83 & $2.0$$\cdot$$10^{-7}$ & $5.8$$\cdot$$10^{-11}$ &   1.9 &  21.09 &   0.83 &   4.82 & $6.4$$\cdot$$10^5$ \\
  18.0 &  5.618 &  0.348 &   5.10 &   4.81 &   1.4 &   4.8 &   5.1 &   2.51 &   0.72 & $2.2$$\cdot$$10^{-8}$ & $6.9$$\cdot$$10^{-11}$ &   1.9 &  18.09 &   0.72 &   2.51 & $6.3$$\cdot$$10^5$ \\
  17.0 &  5.576 &  0.335 &   5.10 &   4.93 &   1.3 &   2.8 &   2.9 &   1.14 &   0.53 & $2.6$$\cdot$$10^{-9}$ & $9.8$$\cdot$$10^{-11}$ &   1.9 &  13.21 &   0.53 &   1.14 & $6.2$$\cdot$$10^5$ \\
  16.0 &  5.531 &  0.320 &   5.10 &   4.98 &   1.2 &   2.0 &   2.1 &   0.65 &   0.40 & $7.7$$\cdot$$10^{-10}$ & $1.4$$\cdot$$10^{-10}$ &    1.9 &   9.86 &   0.40 &   0.65 & $5.9$$\cdot$$10^5$ \\
  15.0 &  5.482 &  0.305 &   5.10 &   5.02 &   1.2 &   1.7 &   1.7 &   0.42 &   0.29 & $3.6$$\cdot$$10^{-10}$ & $2.0$$\cdot$$10^{-10}$ &    2.0 &   7.44 &   0.29 &   0.42 & $5.6$$\cdot$$10^5$ \\
  14.0 &  5.429 &  0.289 &   5.09 &   5.04 &   1.1 &   -- &   1.5 &   --  &   --  &   --  &   --  &   -- &   6.14 &   --  &  --   &   --  \\
  13.0 &  5.371 &  0.273 &   5.09 &   5.05 &   1.1 &   -- &   1.3 &   --  &   --  &   --  &   --  &   -- &   4.45 &   --  &  --   &   --  \\
  12.0 &  5.307 &  0.255 &   5.08 &   5.05 &   1.0 &   -- &   1.2 &   --  &   --  &   --  &   -- &   -- &   3.01 &   --  &  --   &   --  \\
  10.0 &  5.155 &  0.216 &   5.07 &   5.05 &   0.9 &   -- &   1.0 &   --  &   --  &   --  &   -- &   -- &   1.43 &   --  &  --   &   --  \\
   8.0 &  4.959 &  0.172 &   5.05 &   5.04 &   0.8 &   -- &   0.8 &   --  &   --  &   --  &   -- &   -- &   0.69 &   --  &  --   &   --  \\
\hline
\multicolumn{8}{l}{\rule{0cm}{2.2ex} $X=0.00$, $D=10$}\\
\hline
\rule{0cm}{2.2ex}
  15.5 &  5.507 &  0.313 &   5.10 &   4.66 &   1.2 &   7.3 &   8.9 &   5.10 &   0.84 & $1.5$$\cdot$$10^{-7}$ & $6.7$$\cdot$$10^{-11}$ &   2.2 &  23.29 &   0.84 &   5.10 & $5.6$$\cdot$$10^5$ \\
  15.0 &  5.482 &  0.305 &   5.09 &   4.80 &   1.2 &   4.2 &   4.6 &   2.52 &   0.72 & $1.3$$\cdot$$10^{-8}$ & $8.1$$\cdot$$10^{-11}$ &   2.0 &  18.22 &   0.72 &   2.52 & $5.4$$\cdot$$10^5$ \\
  14.0 &  5.429 &  0.289 &   5.09 &   4.92 &   1.1 &   -- &   2.5 &   --  &   --  &   --  &   --  &   -- &  14.52 &   --  &  --   &   --  \\
  13.0 &  5.371 &  0.273 &   5.09 &   4.98 &   1.1 &   -- &   1.8 &   --  &   --  &   --  &   --  &   -- &  10.24 &   --  &  --   &   --  \\
  12.0 &  5.307 &  0.255 &   5.08 &   5.01 &   1.0 &   -- &   1.5 &   --  &   --  &   --  &   -- &   -- &   6.97 &   --  &  --   &   --  \\
  10.0 &  5.155 &  0.216 &   5.07 &   5.03 &   0.9 &   -- &   1.1 &   --  &   --  &   --  &   -- &   -- &   3.07 &   --  &  --   &   --  \\
   8.0 &  4.959 &  0.172 &   5.05 &   5.03 &   0.8 &   -- &   0.9 &   --  &   --  &   --  &   -- &   -- &   1.33 &   --  &  --   &   --  \\
\hline
\multicolumn{8}{l}{\rule{0cm}{2.2ex} $X=0.00$, $D=16$}\\
\hline
\rule{0cm}{2.2ex}
  14.0 &  5.429 &  0.289 &   5.08 &   4.65 &   1.2 &   5.9 &   8.5 &   3.97 &   0.80 & $1.0$$\cdot$$10^{-7}$ & $7.4$$\cdot$$10^{-11}$ &   1.5 &  14.59 &   0.80 &   3.97 & $4.7$$\cdot$$10^5$ \\
  13.5 &  5.401 &  0.281 &   5.08 &   4.80 &   1.2 &   3.5 &   4.3 &   1.97 &   0.66 & $8.1$$\cdot$$10^{-8}$ & $8.9$$\cdot$$10^{-11}$ &   1.4 &  10.99 &   0.66 &   1.97 & $4.0$$\cdot$$10^5$ \\
  13.0 &  5.371 &  0.273 &   4.88 &   4.87 &   2.8 &   -- &   3.0 &   --  &   --  &   --  &   --  &   -- &   0.56 &   --  &  --   &   --  \\
  12.0 &  5.307 &  0.255 &   5.08 &   4.95 &   1.0 &   -- &   1.9 &   --  &   --  &   --  &   --  &   -- &  10.49 &   --  &  --   &   --  \\
  10.0 &  5.155 &  0.216 &   5.07 &   5.01 &   0.9 &   -- &   1.2 &   --  &   --  &   --  &   -- &   -- &   4.75 &   --  &  --   &   --  \\
   8.0 &  4.959 &  0.172 &   5.05 &   5.02 &   0.8 &   -- &   0.9 &   --  &   --  &   --  &   -- &   -- &   2.06 &   --  &  --   &   --  \\
\hline
\end{tabular}}
\tablefoot{\changed Given are: the stellar mass $M_\star$,
  the stellar luminosity $L_\star$,
  the classical Eddington parameter $\Gamma_{\rm e}$ (Eq.\,\ref{eq:gammae}),
  the effective temperature $T_{\rm c}$ related to the core radius $R_{\rm c}$,
  the effective temperature $T_\star$ related to the stellar surface radius $R_\star$,
  the inner core radius $R_{\rm c}$ (at the bottom of the inflated envelope),
  the outer envelope radius $R_{\rm e}$,
  the stellar surface radius $R_\star$,
  the envelope extension $\Delta R$ in units of $R_{\rm c}$, and $R_{\rm e}$,
  the mass $\Delta M$ of the outer shell (with $R_{\rm e} < r < R_\star$),
  the ``minimum density'' $\rho_{\rm min}$ at the point of the lowest gas pressure within the inflated envelope,
  the ratio $f_\rho = \bar{\rho}/\rho_{\rm min}$ (with $\bar{\rho}$ from Eqs.\,\ref{eq:INT}, \ref{eq:wcdef}),
  the ratio $Q$ according to Eq.\,\ref{eq:Q},
  the parameters $W_c$ \& $W_e$ according to Eqs.\,\ref{eq:wcdef}, \ref{eq:wedef},
  and the total pressure $P_{\rm e}$ at the outer envelope radius $R_{\rm e}$.
  The values given in this table are inferred numerically from our computational models,
  and may be compared to the analytical results from Sect.\,\ref{sec:anal}.}
\end{table*}

\begin{table*}
  \caption{Model parameters for chemically-homogeneous stars containing hydrogen. \label{tab:modparH}}
  \centering 
  {\tiny
  \begin{tabular}{rllllllllllllllll} 
    \hline \hline 
    \rule{0cm}{2.8ex} $M_\star$ & $L_\star$ & $\Gamma_{\rm e}$ & $\log(T_{\rm c})$ & $\log(T_\star)$ & $R_{\rm c}$ & $R_{\rm e}$ & $R_\star$ & $\Delta R$ & $\Delta R$ & $\Delta M$ & $\rho_{\rm min}$ & $f_\rho$ & $Q$ & $W_c$ & $W_e$ & $P_{\rm e}$ \\
  $[M_\odot]$ & $[L_\odot]$ & & $[{\rm K}]$ & $[{\rm K}]$ & $[R_\odot]$ & [$R_\odot$] & $[R_\odot]$ & $[R_{\rm c}]$ & $[R_{\rm e}]$ & $[M_\odot]$ & $[\frac{\rm g}{{\rm cm}^3}]$ & & & & & $[\frac{\rm dyn}{{\rm cm}^2}]$ \\
\hline
\multicolumn{8}{l}{\rule{0cm}{2.2ex} $X=0.70$, $D=1$}\\
\hline
\rule{0cm}{2.2ex}
  155.0&  6.418 &  0.435 &   4.70 &   4.20 &   21.3 &  163.3 &  217.6 &   6.68 &   0.87 & $7.1$$\cdot$$10^{-3}$ & $1.0$$\cdot$$10^{-10}$ &   2.3 &  14.27 &   0.87 &   6.68 & $8.6$$\cdot$$10^5$ \\
  150.0&  6.397 &  0.427 &   4.70 &   4.38 &   20.7 &   80.0 &   90.6 &   2.86 &   0.74 & $3.2$$\cdot$$10^{-4}$ & $1.3$$\cdot$$10^{-10}$ &   2.2 &  11.65 &   0.74 &   2.86 & $8.6$$\cdot$$10^5$ \\
  145.0&  6.375 &  0.420 &   4.71 &   4.47 &   19.9 &   53.6 &   58.0 &   1.69 &   0.63 & $6.2$$\cdot$$10^{-5}$ & $1.5$$\cdot$$10^{-10}$ &   2.2 &  10.08 &   0.63 &   1.69 & $8.6$$\cdot$$10^5$ \\
  140.0&  6.351 &  0.412 &   4.71 &   4.53 &   19.2 &   41.1 &   43.6 &   1.14 &   0.53 & $2.2$$\cdot$$10^{-5}$ & $1.9$$\cdot$$10^{-10}$ &   2.2 &   8.62 &   0.53 &   1.14 & $8.5$$\cdot$$10^5$ \\
  130.0&  6.301 &  0.396 &   4.71 &   4.60 &   17.9 &   29.2 &   30.2 &   0.63 &   0.39 & $5.5$$\cdot$$10^{-6}$ & $2.9$$\cdot$$10^{-10}$ &   2.3 &   6.45 &   0.39 &   0.63 & $8.0$$\cdot$$10^5$ \\
  120.0&  6.247 &  0.378 &   4.71 &   4.63 &   16.7 &   -- &   23.9 &   --  &   --  &   --  &   --  &   -- &   5.46 &   --  &  --   &   --  \\
  110.0&  6.186 &  0.359 &   4.71 &   4.66 &   15.7 &   -- &   20.2 &   --  &   --  &   --  &   --  &   -- &   3.77 &   --  &  --   &   --  \\
  100.0&  6.118 &  0.337 &   4.71 &   4.67 &   14.6 &   -- &   17.6 &   --  &   --  &   --  &   -- &   -- &   2.58 &   --  &  --   &   --  \\
   90.0&  6.041 &  0.314 &   4.70 &   4.67 &   13.7 &   -- &   15.7 &   --  &   --  &   --  &   -- &   -- &   1.76 &   --  &  --   &   --  \\
   80.0&  5.952 &  0.288 &   4.70 &   4.67 &   12.7 &   -- &   14.1 &   --  &   --  &   --  &   -- &   -- &   1.20 &   --  &  --   &   --  \\
   70.0&  5.848 &  0.259 &   4.69 &   4.67 &   11.7 &   -- &   12.7 &   --  &   --  &   --  &   -- &   -- &   0.81 &   --  &  --   &   --  \\
   60.0&  5.724 &  0.227 &   4.68 &   4.66 &   10.7 &   -- &   11.5 &   --  &   --  &   --  &   --  &   -- &   0.55 &   --  &  --   &   --  \\
   50.0&  5.569 &  0.191 &   4.66 &   4.65 &   9.6 &   -- &   10.2 &   --  &   --  &   --  &   --  &   -- &   0.36 &   --  &  --   &   --  \\
   40.0&  5.368 &  0.150 &   4.63 &   4.63 &   8.9 &   -- &   8.9 &   --  &   --  &   --  &   --  &   -- &   0.10 &   --  &  --   &   --  \\
   30.0&  5.087 &  0.105 &   4.59 &   4.59 &   7.6 &   -- &   7.6 &   --  &   --  &   --  &   -- &   -- &   0.09 &   --  &  --   &   --  \\
   20.0&  4.647 &  0.057 &   4.53 &   4.53 &   6.0 &   -- &   6.1 &   --  &   --  &   --  &   -- &   -- &   0.09 &   --  &  --   &   --  \\
\hline
\multicolumn{8}{l}{\rule{0cm}{2.2ex} $X=0.40$, $D=1$}\\
\hline
\rule{0cm}{2.2ex}
   80.0&  6.179 &  0.400 &   4.73 &   4.29 &   14.3 &   86.8 &  106.6 &   5.07 &   0.84 & $8.9$$\cdot$$10^{-4}$ & $1.6$$\cdot$$10^{-10}$ &   2.6 &  12.21 &   0.84 &   5.07 & $8.1$$\cdot$$10^5$ \\
   75.0&  6.135 &  0.385 &   4.73 &   4.50 &   13.4 &   35.7 &   38.4 &   1.66 &   0.62 & $2.2$$\cdot$$10^{-5}$ & $2.3$$\cdot$$10^{-10}$ &   2.7 &   9.46 &   0.62 &   1.66 & $7.7$$\cdot$$10^5$ \\
   70.0&  6.088 &  0.370 &   4.72 &   4.59 &   13.1 &   23.8 &   24.8 &   0.82 &   0.45 & $4.3$$\cdot$$10^{-6}$ & $3.4$$\cdot$$10^{-10}$ &   2.3 &   5.43 &   0.45 &   0.82 & $7.6$$\cdot$$10^5$ \\
   65.0&  6.036 &  0.354 &   4.73 &   4.63 &   12.3 &   -- &   19.0 &   --  &   --  &   --  &   --  &   -- &   4.73 &   --  &  --   &   --  \\
   60.0&  5.979 &  0.336 &   4.73 &   4.66 &   11.5 &   -- &   15.7 &   --  &   --  &   --  &   -- &   -- &   3.38 &   --  &  --   &   --  \\
   50.0&  5.844 &  0.296 &   4.72 &   4.68 &   10.1 &   -- &   12.0 &   --  &   --  &   --  &   -- &   -- &   1.69 &   --  &  --   &   --  \\
   40.0&  5.671 &  0.248 &   4.71 &   4.68 &   8.8 &   -- &   9.8 &   --  &   --  &   --  &   -- &   -- &   0.83 &   --  &  --   &   --  \\
   30.0&  5.431 &  0.190 &   4.68 &   4.67 &   7.4 &   -- &   8.0 &   --  &   --  &   --  &   -- &   -- &   0.39 &   --  &  --   &   --  \\
   20.0&  5.054 &  0.120 &   4.63 &   4.63 &   6.2 &   -- &   6.2 &   --  &   --  &   --  &   -- &   -- &   0.09 &   --  &  --   &   --  \\
   15.0&  4.756 &  0.080 &   4.59 &   4.59 &   5.3 &   -- &   5.3 &   --  &   --  &   --  &   -- &   -- &   0.09 &   --  &  --   &   --  \\
\hline
\multicolumn{8}{l}{\rule{0cm}{2.2ex} $X=0.20$, $D=1$}\\
\hline
\rule{0cm}{2.2ex}
   46.0&  5.975 &  0.372 &   4.74 &   4.37 &   10.7 &   50.9 &   59.5 &   3.75 &   0.79 & $1.6$$\cdot$$10^{-4}$ & $2.3$$\cdot$$10^{-10}$ &   2.4 &   7.97 &   0.79 &   3.75 & $7.8$$\cdot$$10^5$ \\
   45.0&  5.960 &  0.367 &   4.74 &   4.45 &   10.7 &   36.7 &   40.8 &   2.43 &   0.71 & $4.1$$\cdot$$10^{-5}$ & $2.6$$\cdot$$10^{-10}$ &   2.2 &   6.56 &   0.71 &   2.43 & $7.7$$\cdot$$10^5$ \\
   44.0&  5.944 &  0.362 &   4.74 &   4.50 &   10.4 &   29.0 &   31.5 &   1.78 &   0.64 & $1.6$$\cdot$$10^{-5}$ & $2.9$$\cdot$$10^{-10}$ &   2.2 &   6.00 &   0.64 &   1.78 & $7.7$$\cdot$$10^5$ \\
   42.0&  5.911 &  0.352 &   4.74 &   4.57 &   10.0 &   20.9 &   22.1 &   1.09 &   0.52 & $4.2$$\cdot$$10^{-6}$ & $3.8$$\cdot$$10^{-10}$ &   2.2 &   4.86 &   0.52 &   1.09 & $7.5$$\cdot$$10^5$ \\
   40.0&  5.877 &  0.341 &   4.74 &   4.61 &    9.6 &   16.8 &   17.4 &   0.75 &   0.43 & $1.7$$\cdot$$10^{-6}$ & $4.9$$\cdot$$10^{-10}$ &   2.3 &   4.10 &   0.43 &   0.75 & $7.0$$\cdot$$10^5$ \\
   35.0&  5.780 &  0.312 &   4.74 &   4.67 &   8.6 &   -- &   11.9 &   --  &   --  &   --  &   --  &   -- &   2.73 &   --  &  --   &   --  \\
   30.0&  5.664 &  0.279 &   4.73 &   4.69 &   7.7 &   -- &   9.4 &   --  &   --  &   --  &   --  &   -- &   1.53 &   --  &  --   &   --  \\
   25.0&  5.520 &  0.240 &   4.72 &   4.70 &   6.8 &   -- &   7.8 &   --  &   --  &   --  &   -- &   -- &   0.84 &   --  &  --   &   --  \\
   20.0&  5.334 &  0.195 &   4.71 &   4.69 &   6.0 &   -- &   6.5 &   --  &   --  &   --  &   -- &   -- &   0.45 &   --  &  --   &   --  \\
   15.0&  5.073 &  0.143 &   4.68 &   4.66 &   5.1 &   -- &   5.4 &   --  &   --  &   --  &   -- &   -- &   0.23 &   --  &  --   &   --  \\
   10.0&  4.664 &  0.084 &   4.61 &   4.61 &   4.3 &   -- &   4.3 &   --  &   --  &   --  &   -- &   -- &   0.09 &   --  &  --   &   --  \\
\hline
\multicolumn{8}{l}{\rule{0cm}{2.2ex} $X=0.36$, $D=1$ (AG Car)}\\
\hline
\rule{0cm}{2.2ex}
   73.0&  6.149 &  0.397 &   4.72 &   4.23 &   14.3 &  104.3 &  136.8 &   6.30 &   0.86 & $2.2$$\cdot$$10^{-3}$ & $1.6$$\cdot$$10^{-10}$ &   2.2 &   9.68 &   0.86 &   6.30 & $8.0$$\cdot$$10^5$ \\
   70.0&  6.120 &  0.387 &   4.72 &   4.44 &   13.8 &   46.1 &   51.3 &   2.36 &   0.70 & $6.8$$\cdot$$10^{-5}$ & $2.0$$\cdot$$10^{-10}$ &   2.2 &   7.69 &   0.70 &   2.36 & $8.0$$\cdot$$10^5$ \\
   65.0&  6.069 &  0.371 &   4.73 &   4.56 &   12.7 &   25.6 &   26.9 &   1.02 &   0.50 & $6.3$$\cdot$$10^{-6}$ & $3.1$$\cdot$$10^{-10}$ &   2.3 &   5.89 &   0.50 &   1.02 & $7.7$$\cdot$$10^5$ \\
   60.0&  6.013 &  0.353 &   4.73 &   4.62 &   11.8 &   -- &   19.2 &   --  &   --  &   --  &   --  &   -- &   4.90 &   --  &  --   &   --  \\
\hline
\end{tabular}}
\tablefoot{\changed See Table\,\ref{tab:modparHE} for an explanation of the listed physical quantities.}
\end{table*}

\begin{figure}[t!]
\parbox[b]{0.49\textwidth}{\includegraphics[scale=0.45]{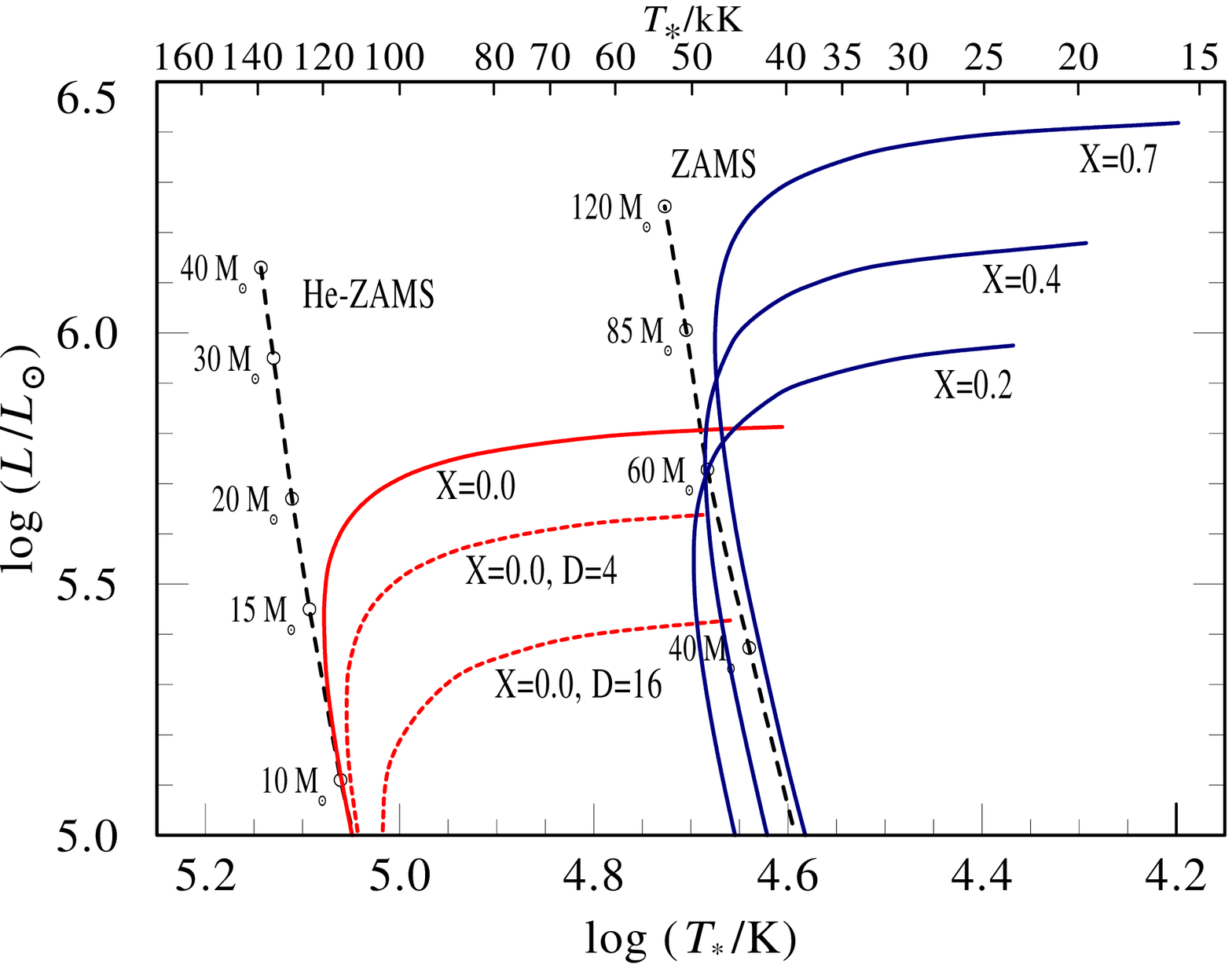}}  
\caption{HR diagram for pure helium models (red), and models
  containing hydrogen (blue).}
  \label{fig:hrdmod}
\end{figure}

To assess the effect of the envelope inflation in the HR diagram, we
have computed a grid of homogeneous stellar structure models with
hydrogen mass fractions $X = 0.0$, 0.2, 0.4, and 0.7, that covers a
luminosity range between $\log(L/L_\odot) = 5.0$, and 6.5. The results
are compiled in Tables\,\ref{tab:modparHE} and \ref{tab:modparH}.  In
Fig.\,\ref{fig:hrdmod}, we show the position of our models in the HR
diagram, together with theoretical zero-age main sequences (ZAMS) that
do not take the inflation effect into account.

At low luminosities, our He-burning models (with $X=0.0$) closely
resemble the He-ZAMS from \citet{lan1:89}, and our H-burning models
($X = 0.2$--0.7) the ZAMS from \citet{sch1:92}. Towards higher
luminosities, our models develop inflated envelopes, i.e., they show
lower $T_\star$. With increasing luminosity, $T_\star$ decreases
rapidly until, above a certain luminosity, we do not find static
solutions anymore. The reason for the occurrence of this important
limit will be explained in Sect.\,\ref{sec:rcalc}.

In our present models, the envelope inflation sets in for Eddington
factors $\Gamma_{\rm e}$ between 0.3 ($X=0.0$), and 0.4 ($X=0.7$),
i.e., within a relatively narrow range. The fact that it sets in
earlier for H-deficient stars is a consequence of the fact that the
$L/M$ ratios of these objects are higher (although $\kappa_{\rm e}$ is
lower for lower H abundances).  In Sects.\,\ref{sec:rcalc}, \&
\ref{sec:recipe} we will show that this behavior, as well as the
described upper luminosity limit, are a consequence of the strength,
and shape of the Fe-opacity peak in the $P_{\rm rad}$--$P_{\rm gas}$
plane.

\subsection{The influence of density inhomogeneities}
\label{sec:clumping}

For our pure He models in Table\,\ref{tab:modparHE}, the described
envelope extension starts at luminosities of
$\log(L/L_\odot)\approx5.5$, and fully develops for
$\log(L/L_\odot)\approx5.8$ (cf.\ Fig.\,\ref{fig:hrdmod}).  As we will
discuss in Sect.\,\ref{sec:wr}, the largest part of the Galactic
H-free Wolf-Rayet stars, however, seems to show inflated envelopes for
luminosities just below this value. In the following we will show that
the inflation in this parameter range can be explained by density
inhomogeneities. Such inhomogeneities are expected to arise in the
outer envelopes of WR stars, e.g.\ due to strange-mode instabilities
\citep{gla1:02,gla1:08}.

{\changed 

  If material is inhomogeneous, or clumped, the density within clumps
  is larger than the mean density. As a result, the opacity $\kappa$
  has to be evaluated for an increased density, instead of the mean
  density (note that $\kappa$ denotes the mass absorption coefficient
  in ${\rm cm}^2/{\rm g}$). Here we assume a constant clumping factor
  $D$ within the inflated envelope, and evaluate $\kappa(\rho,T)$ for
  the enhanced density $D\times\rho$, instead of the mean density
  $\rho$.  This is equivalent to assuming a medium with clumps of a
  constant density $D\times\rho$, and an inter-clump medium which is
  void. In this picture the clumps would thus only fill a fraction
  $f=1/D$ of the total volume. The same definition of clumping factors
  $D$, and filling factors $f$ is commonly used for the modeling of
  inhomogeneous stellar winds \citep[e.g.][]{ham1:98}.

  In the clumping approximation used here, it is assumed that no
  radiative flux gets lost in between clumps, e.g.\ by
  shadowing/porosity effects. Such a situation could be achieved by
  optically thin clumps, or by a clump geometry (such as large shells,
  or pancakes) that is not sensitive to shadowing effects.

}

Following our argumentation from the previous section, large envelope
extensions, i.e., large scale heights $H$, are reached for Eddington
factors $\Gamma$ that are very close to one. We have shown that this
is achieved for the low values of $P_{\rm gas}$ at the tip of the
Fe-opacity peak, as $P_{\rm gas}/P \approx (1-\Gamma)$. {\changed By
  the assumption of clumping, the opacity peak is shifted towards even
  lower {\em mean densities}, i.e., towards lower (mean) values of
  $P_{\rm gas}$. In this way, the envelope inflation is further
  enhanced} (a more concise explanation of this effect is given in
Sect.\,\ref{sec:anal}, \,Eqs.\,\ref{eq:wcdef},
\ref{eq:WCWE}\, \ref{eq:W}).

In Fig.\,\ref{fig:hrdmod} we show He-ZAMS models that are computed
with clumping factors $D=4$, and $D=16$.  As expected, the envelope
extension occurs much earlier in these models. In Sect.\,\ref{sec:wr}
we will show that these models cover the observed HR diagram (HRD)
positions of the Galactic H-free WR stars, i.e., we can explain the
observed temperatures of these objects with moderate clumping factors.
The adopted values {\changed for $D$} are in notable agreement with
spectroscopically determined clumping factors for the winds of WR
stars, {\changed e.g.\ by \citet{ham1:98}.}

\section{Analytical description of the envelope inflation}
\label{sec:anal}

In the previous section we have shown that the properties of inflated
envelopes are chiefly determined by the fact that the solution has to
follow a path with $\Gamma = 1$ in the $P_{\rm rad}$--$P_{\rm gas}$
plane (cf.\ Fig.\,\ref{fig:gamma}). In the following we elaborate on
this, and derive an analytical description of this process. {\changedA
  As we concentrate on the physics of the inflated envelopes {\em
    alone}, the resulting relations do not depend on the internal
  structure of the star, i.e., they are generally applicable to stars
  with given (observed) parameters.} In Sect.\,\ref{sec:anainf}, we
start with the equations describing the envelope structure. In
Sect.\,\ref{sec:rcalc} we derive analytical expressions for the radial
extension of the envelope $\Delta R$, and in Sect.\,\ref{sec:recipe}
we present a recipe to estimate $\Delta R$ based on observed/adopted
stellar parameters.  Furthermore, we derive an estimate for the mass
of the surrounding shell, $\Delta M$, in Sect.\,\ref{sec:mcalc}.
Finally, we discuss {\changed in Sect.\,\ref{sec:assumptions}, to
  which extent the inflation effect may be affected by model
  assumptions.}

\subsection{Inflated low-density envelopes near the Eddington limit}
\label{sec:anainf}

In this section we concentrate on the physics of the outermost stellar
envelope, where $m = M$, and $L=const.$ With these assumptions, we
eliminate two equations from the stellar structure
equations\,(Eqs.\,\ref{eq:hydeq}--\ref{eq:etrans}), so that only the
equation of hydrostatic equilibrium (Eq.\,\ref{eq:hydeq}), and the
energy transport equation (Eq.\,\ref{eq:etrans}) are left.

In Sect.\,\ref{sec:inflation} we concluded, that two conditions are
mandatory to maintain an inflated envelope.  1) the star needs to be
close enough to the Eddington limit, so that $\Gamma = 1$ can be
reached, and 2) convective energy transport needs to be inefficient.
Under condition 2) (the validity of this condition is discussed in
Sect.\,\ref{sec:conveff}), the energy transport equation
(Eq.\,\ref{eq:etrans}) becomes purely radiative, i.e., $L=L_{\rm
  rad}$, and $\nabla = \nabla_{\rm rad}$, with
\begin{equation}
  \nabla_{\rm rad} = \frac{3}{16\pi a c}\frac{P\kappa}{T^4}\frac{L}{GM}.
\end{equation}
After substituting Eq.\,\ref{eq:hydeq}, and Eq.\,\ref{eq:mass} in
Eq.\,\ref{eq:etrans}, we thus obtain the energy transport equation
in the form
\begin{equation} 
  \frac{\partial T}{\partial r} =
  -\frac{3}{4 a T^3}\frac{\kappa \rho L}{4 \pi r^2 c}.
\end{equation}
If this relation is written in terms of $P_{\rm rad}=(a/3)T^4$ we
obtain
\begin{equation} 
  \frac{\partial P_{\rm rad}}{\partial r} =
  \frac{4 a T^3}{3} \frac{\partial T}{ \partial r} = 
  - \frac{\kappa \rho L}{4 \pi r^2 c} = - \rho g_{\rm rad},
\label{eq:diff}
\end{equation}
for the (outward directed) radiative acceleration $g_{\rm rad}$.
Together with the equation of hydrostatic equilibrium
\begin{equation} 
  \frac{\partial P}{\partial r} =
  \frac{\partial (P_{\rm gas}+P_{\rm rad})}{\partial r} =
  -\frac{GM\rho}{r^2} = -\rho g,
\label{eq:HSE}
\end{equation}
we obtain by division of Eq.\,\ref{eq:HSE} and Eq.\,\ref{eq:diff}
\begin{equation}
  \frac{\partial P}{\partial P_{\rm rad}} =
  \frac{\partial P_{\rm gas}}{\partial P_{\rm rad}} + 1 =
  \frac{g}{g_{\rm rad}}= \frac{1}{\Gamma},
\label{eq:GammaEQ}
\end{equation}
or
\begin{equation} 
  \frac{\partial P_{\rm gas}}{\partial P_{\rm rad}} = \frac{1}{\Gamma}-1.
\label{eq:stan}
\end{equation}

Notably, ${\partial P_{\rm gas}}/{\partial P_{\rm rad}}$ depends {\em
  only} on the Eddington factor $\Gamma$, i.e., via
$\Gamma/\Gamma_{\rm e} = \kappa/\kappa_{\rm e}$ on $\Gamma_{\rm e}$,
and $\kappa$.
This is in line with our previous finding that the envelope inflation
is fully determined by the classical Eddington factor $\Gamma_{\rm e}$
(as given by Eq.\,\ref{eq:gammae}), which is a function of stellar
parameters, and $\kappa$, which depends on material properties.

In Fig.\,\ref{fig:gamma} we compare our numerical solution from
Sect.\,\ref{sec:inflation} (blue line) with the slopes ${\partial
  P_{\rm rad}}/{\partial P_{\rm gas}}$ (black arrows) following from
Eq.\,\ref{eq:stan}, in the $P_{\rm rad}$--$P_{\rm gas}$ plane. The
background colors indicate the value of $\Gamma = \Gamma_{\rm e}
(\kappa/\kappa_{\rm e})$ for our model, where $\kappa$ has been
extracted from the OPAL opacity tables \citep{igl1:96}.

According to Eq.\,\ref{eq:stan}, the point at which the envelope
solution crosses the Eddington limit, i.e. where $\Gamma = 1$, needs
to be an extremum in $P_{\rm gas}$.  Fig.\,\ref{fig:gamma} nicely
shows, that our numerical solution indeed approaches the Eddington
limit, and crosses it at the point of the lowest possible gas
pressure. If the Eddington limit would be crossed earlier, at a higher
density, the gas pressure had to increase so fast, that extremely high
densities would be required at the outer boundary.  The only way to
reach reasonably low temperatures, and densities at the outer
boundary, thus leads {\em around} the Fe-opacity peak, via a low gas
pressure, i.e., via low densities. The formation of a low-density
envelope with a density inversion is thus a natural consequence of the
topology of the Fe opacity peak in the $P_{\rm rad}$--$P_{\rm gas}$
plane, in combination with Eq.\,\ref{eq:stan}.

For the example in Fig.\,\ref{fig:gamma}, we find that $P_{\rm
  gas}/P_{\rm rad} \approx 4$$\cdot$$10^{-4}$ at the tip of the Fe-opacity
peak. In Sect.\,\ref{sec:inflation} we already discussed that this
implies that $\Gamma \rightarrow 1$, as $P_{\rm gas}/P \approx
(1-\Gamma)$. This relation follows from Eq.\,\ref{eq:GammaEQ}, if we
assume a slowly changing ratio $P_{\rm rad}/P$.  In this case we have
$\partial (\ln P_{\rm rad})/\partial (\ln P) \approx 1$ (cf.\,
Fig.\,\ref{fig:pgasprad}). It thus follows that
\begin{equation} 
  \frac{\partial (\ln P_{\rm rad})}{\partial (\ln P)} = 
  \frac{P}{P_{\rm rad}} \frac{\partial P_{\rm rad}}{\partial P} =
  \frac{P}{P_{\rm rad}} \, \Gamma \approx 1,
\end{equation}
which leads to
\begin{equation} 
  \frac{P_{\rm rad}}{P} \approx \Gamma,
\end{equation}
or
\begin{equation} 
  \frac{P_{\rm gas}}{P} = \frac{P_{\rm gas}}{P_{\rm gas}+P_{\rm rad}} \approx (1-\Gamma).
  \label{eq:PGAM}
\end{equation}
As discussed in Sect.\,\ref{sec:inflation}, this small value of
$(1-\Gamma)$ leads to a density scale height of the order of the
stellar radius, i.e., to an envelope inflation.

The occurrence of an envelope inflation is thus intimately connected
to the fact that $P_{\rm gas} \ll P_{\rm rad}$ at the tip of the
Fe-opacity peak. As noted earlier, the location of this point in the
${P_{\rm rad}}$--${P_{\rm gas}}$ plane is only determined by the
topology of the opacity peak, i.e., by material properties. In the
following we will take advantage of this fact, to obtain analytical
estimates of the radial envelope extension $\Delta R$.

\subsection{Computation of the radial extension $\Delta R$}
\label{sec:rcalc}

To compute the radial extension $\Delta R$, we integrate
Eq.\,\ref{eq:diff} and Eq.\,\ref{eq:HSE} from the outer boundary of
the inflated zone, at the ``envelope radius'' $R_{\rm e}$, to the
inner boundary, at the ``core radius'' $R_{\rm c}$.  As this
integration is performed within the inflated zone, where $P=P_{\rm
  rad}$, and $\kappa = \kappa_{\rm Edd}$, Eqs.\,\ref{eq:diff} and
\ref{eq:HSE} degenerate to one equation, that describes the change of
$P_{\rm rad}$, and thus also the change of the temperature, in
hydrostatic equilibrium.
\begin{equation} 
  \frac{\partial P_{\rm rad}}{\partial r} =
  -\frac{\kappa_{\rm Edd} \rho L}{4 \pi r^2 c} =
  -\frac{GM\rho}{r^2},
\label{eq:ONE}
\end{equation}
or
\begin{equation} 
  \frac{{\rm d} P_{\rm rad}}{\rho} = GM\, {\rm d}\frac{1}{r}.
\label{eq:DINT}
\end{equation}

The integration is performed from $R_{\rm e}$, where $P\equiv P_{\rm
  e}$, to $R_{\rm c}$, where $P\equiv P_{\rm c}$, leading to
\begin{equation} 
  \frac{R_{\rm c}}{GM}\int_{P_{\rm e}}^{P_{\rm c}}\frac{{\rm d} P_{\rm rad}}{\rho}
  = 1-\frac{R_{\rm c}}{R_{\rm e}}.
\label{eq:INT}
\end{equation}
The left hand side of this equation can be written in the form
\begin{equation} 
W_{\rm c} \equiv \frac{R_{\rm c}}{GM} \frac{\Delta P}{\bar{\rho}},
\label{eq:wcdef}
\end{equation}
where $\Delta P = P_{\rm c} - P_{\rm e}$, and $1/\bar{\rho}$ is the
mean of $1/\rho$ with respect to $P_{\rm rad}$.
 In Sect.\,\ref{sec:recipe} we will show
that $\Delta P$, and $\bar{\rho}$ can be estimated {\em alone} on the
basis of opacity tables. For given $W_{\rm c}$, the radial envelope
extension can then be computed from Eq.\,\ref{eq:INT}, which becomes
\begin{equation}
\frac{R_e}{R_c} = \frac{1}{1-W_{\rm c}} = 1 + W_{\rm e},
\label{eq:WCWE}
\end{equation}
where the latter equality uses a definition of $W$ in terms of the
observable, outer-envelope radius $R_{\rm e}$
\begin{equation} 
  W_{\rm e} \equiv \frac{R_{\rm e}}{GM} \frac{\Delta P}{\bar{\rho}}.
  \label{eq:wedef}
\end{equation}
The expression for the radial extension $\Delta R = R_{\rm e}-R_{\rm
  c}$ for given $W_{\rm c}$, or $W_{\rm e}$ follows directly from
Eq.\,\ref{eq:WCWE}
\begin{equation} 
  \frac{\Delta R}{R_{\rm c}} =
  \frac{1}{1 -  W_{\rm c}} -1,
\label{eq:DeltaRc}
\end{equation}
\begin{equation} 
  \frac{\Delta R}{R_{\rm e}} =
  1-\frac{1}{1 + W_{\rm e}}.
\label{eq:DeltaRe}
\end{equation}

For the $W_{\rm c}$, and $W_{\rm e}$ as determined from our models
(Tables\,\ref{tab:modparHE} and \ref{tab:modparH}) these relations are
fulfilled very precisely.

{\changed The parameters $W_{\rm c/e}$ are of essential importance for
  the inflation effect. As $u=3P/\rho$ in a radiation dominated gas
  (where $u$ denotes the specific energy per volume), $\Delta
  P/\bar{\rho}$ is related to the specific energy per gram of material
  within the inflated zone. According to Eq.\,\ref{eq:INT}, it
  denotes the increase of the specific energy, from the outer to the
  inner edge of the inflated envelope, divided by 3. The parameters
  $W_{\rm c/e}$ indicate the ratio between this energy increase, and
  the gravitational energy per gram of material, at $R_{\rm c/e}$.}

Notably, Eq.\,\ref{eq:WCWE} imposes an instability limit for $W_{\rm
  c}\rightarrow1$, for which $R_{\rm e}\rightarrow \infty$.  This
limit coincides with the limit that we have described in
Sect.\,\ref{sec:grid}, where we found that no hydrostatic solutions
exist, if certain luminosities are exceeded.  Stable solutions only
exist for $W_{\rm c}<1$, which is equivalent to
\begin{equation} 
  \frac{\Delta P}{\bar{\rho}} < \frac{GM}{R_{\rm c}}.
\label{eq:goetz-limit}
\end{equation}

As $\Delta P/\bar{\rho}$ is largely
fixed by material properties, Eq.\,\ref{eq:goetz-limit} imposes a
limit on the gravitational energy at the core radius $R_{\rm c}$. If
the gravitational energy becomes too small (i.e., $R_{\rm c}$ becomes
too large), the stellar envelope becomes energetically unbound, and
there exists no hydrostatic envelope solution.  This limit might be of
profound importance for the occurrence of instabilities close to the
Eddington limit, as observed, e.g.\ for LBVs (cf.\
Sec.\,\ref{sec:lbv}).

The occurrence of the envelope inflation is thus mainly due to the
fact that the Fe-opacity peak forces the envelope to low densities
$\rho$, while the temperature of the opacity peak is fixed, i.e.,
$u=const$. This leads to a very high specific energy $u/\rho$, i.e.,
the envelope becomes less gravitationally bound.

\subsection{A recipe to compute $\Delta R$}
\label{sec:recipe}

\begin{figure}[t!]
  \parbox[b]{0.49\textwidth}{{\includegraphics[scale=0.445]{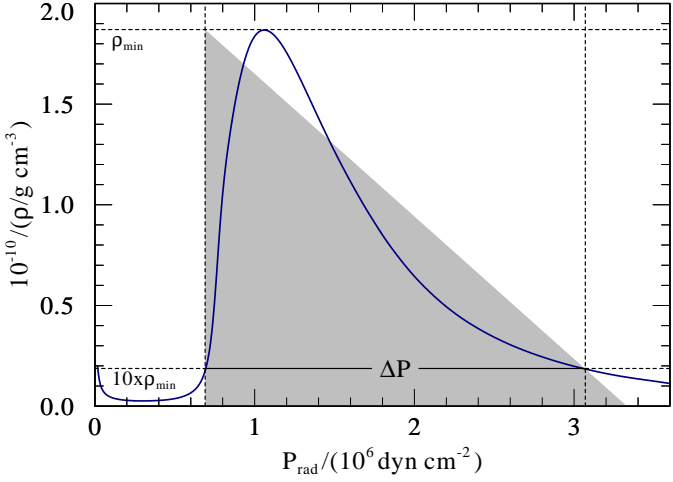}}}
  \caption{Plot of $1/\rho$ vs.\ $P_{\rm rad}$. The dashed horizontal
    lines indicate $1/\rho_{\rm min}$, and $1/(10\,\rho_{\rm min})$.
    The intersections of the latter with the envelope solution
    represent boundary pressure $P_{\rm c}$, and $P_{\rm e}$, as
    indicated by vertical dashed lines. The grey shaded area
    illustrates that the integral $\int \dint P/\rho \approx \Delta
    P/(2\rho_{\rm min})$.}
  \label{fig:drho}
\end{figure}

The remaining step for the computation of $\Delta R$ is to obtain
estimates of the parameter W (Eqs.\,\ref{eq:wcdef}, \ref{eq:wedef}),
which depends on the ratio $M/R$ (where $R$ denotes either $R_{\rm
  c}$, or $R_{\rm e}$), and the ratio $\Delta P/\bar{\rho}$. While
$M$, and $R$ are given stellar parameters, $\Delta P/\bar{\rho}$ needs
to be determined from opacity tables.

Based on Eqs.\,\ref{eq:INT}, \ref{eq:wcdef} we have
\begin{equation} 
\frac{\Delta P}{\bar{\rho}}
= \int_{P_{\rm e}}^{P_{\rm c}}\frac{{\rm d} P_{\rm rad}}{\rho}.
\end{equation}
In Fig.\,\ref{fig:drho} we illustrate that, due to the topology of the
Fe-opacity peak, the integral on the right hand side is roughly equal
to $\Delta P/(2 \rho_{\rm min})$, i.e.,
\begin{equation} 
  W = \frac{R}{GM} \frac{\Delta P}{\bar{\rho}} 
  = \frac{R}{GM} \frac{\Delta P}{f_{\rho}\rho_{\rm min}},
\label{eq:WW}
\end{equation}
with $f_\rho\approx 2$.  Numerically obtained values for $f_\rho$ from
our model computations confirm this value (cf.\ Table
\ref{tab:modparHE} and \ref{tab:modparH}). $W$ can thus be computed on
the basis of $\Delta P = P_{\rm c}-P_{\rm e}$, and $\rho_{\rm min}$.

To extract these values from the opacity tables, we make use the fact
that the envelope solution follows almost precisely a path with
$\Gamma = 1$, i.e., a path with $\kappa = \kappa_{\rm Edd}$ in the
$P_{\rm rad}$--$P_{\rm gas}$ plane. From this path we extract the
minimum {\em gas pressure} $P_{\rm min}$, at the tip of the Fe-opacity
peak (cf.  Fig.\,\ref{fig:gamma}). We use the density at this point as
our reference density $\rho_{\rm min}$. For given $P_{\rm min}$, we
{\em define} the boundaries of the inflated envelope as the two
adjacent points on the path with $\kappa = \kappa_{\rm Edd}$, where
$P_{\rm gas}=f_P\cdot P_{\rm min}$ (in this work we use $f_P=10$). The
{\em radiation pressure} at these two points gives us $P_{\rm c}$, and
$P_{\rm e}$. Our choice of $f_P=10$ has a moderate influence on the
resulting values, will however not change the overall results.

The procedure described above thus provides $\rho_{\rm min}$, and
$\Delta P = P_{\rm c}-P_{\rm e}$ for given $\kappa_{\rm Edd}$, or
$\Gamma_{\rm e}$ (note that $\Gamma_{\rm e} (\kappa_{\rm
  Edd}/\kappa_{\rm e}) = 1$), i.e., we can now compute $W$, and
$\Delta R$. To determine $\Delta P/\rho_{\rm min}$ over a wide
parameter range, we employ a root finding algorithm to extract $\Delta
P$, and $\rho_{\rm min}$ from the OPAL tables \citep{igl1:96}, as
described above. The value of $\Delta P/\rho_{\rm min}$ depends on the
chemical composition, i.e., $X$ and $Z$, and on $\Gamma_{\rm e}$.

\begin{figure}[t!]
\parbox[b]{0.49\textwidth}{\includegraphics[scale=0.45]{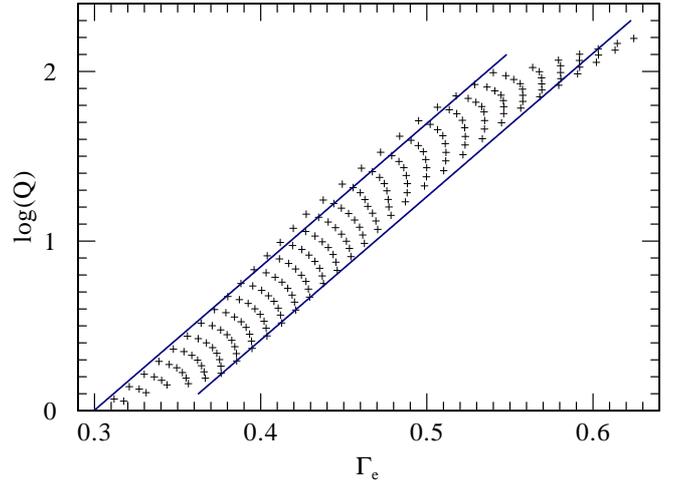}}  
\caption{Fitting relation for $Q(X, Z=Z_\odot, \Gamma_{\rm e})$.
  Black crosses indicate numerically determined values of $Q$, for
  $X=0$ -- 0.7. Blue lines indicate our fitting relation
  Eq.\,\ref{eq:Qrecipe}, for $X=0$ (top), and $X=0.7$ (bottom).}
  \label{fig:Q}
\end{figure}

At this point we define a function $Q(X, Z, \Gamma_{\rm e})$, that
only depends on $\Delta P/\rho_{\rm min}$
\begin{equation}
  Q(X, Z, \Gamma_{\rm e}) =
  \frac{R_\odot \Delta P}{GM_\odot \rho_{\rm min}}.
  \label{eq:Q}
\end{equation}
Adopting a Galactic metallicity of $Z=0.02$, we determine a fitting
relation for $Q(X, \Gamma_{\rm e})$, based on the numerically
determined values of $\Delta P/\rho_{\rm min}$. The resulting relation
has the form 
\begin{equation}
  \log(Q) = -2.135 - 0.617\,X + 8.453\,\Gamma_{\rm e}.
  \label{eq:Qrecipe}
\end{equation} 
The fits are displayed in Fig.\,\ref{fig:Q}, where the crosses denote
the numerically determined values of $Q$, for the range $X=0$--0.7,
and the blue lines denote the fitting relation
Eq.\,\ref{eq:Qrecipe}, for $X=0$, and $X=0.7$.

The values of $Q$ determined this way, lie between $Q\sim 1$, and
$Q\sim 150$ (cf.\ Fig.\,\ref{fig:Q}), with $\Gamma_{\rm e} =
0.3$--0.6.  For lower $Q$ (i.e., lower $\Gamma_{\rm e}$), the
extension of the Fe-peak becomes so small that it does not cover a
factor $f_P=10$ in $P_{\rm gas}$.  Under these conditions an envelope
inflation may still occur, however only of moderate strength.  For $Q$
larger than $\sim 150$, $\rho_{\rm min}$ becomes so small that the
range of the opacity tables is exceeded. In this case we still expect
a very strong envelope inflation, but are just not able to infer $Q$
from the tables.

For given $Q$, $W$ can be determined from Eq.\,\ref{eq:WW}. If we
additionally allow for clumping with a clumping factor $D$
(cf.\,Sect.\,\ref{sec:clumping}), we have $\bar{\rho}= (f_\rho/D)
\rho_{\rm min}$, i.e., we can express $W$ in the form
\begin{equation}
  W = 
  \frac{R\, \Delta P}{GM \rho_{\rm min} (f_\rho / D)} =
  \frac{Q(X, Z, \Gamma_{\rm e})}{(f_\rho/D)} \frac{(R/R_\odot)}{(M/M_\odot)},
\label{eq:W}
\end{equation}
with $f_\rho\approx 2$.

To compute the envelope extension $\Delta R$ for a star with given
parameters $M$, $L$, $X$, and $R$ (where $R$ may denote $R_{\rm c}$ or
$R_{\rm e}$), the following steps thus need to be performed.
\begin{itemize}
\item[1)] For given $M$, $L$, and $X$, $\Gamma_{\rm e}$ can be
  computed from Eq.\,\ref{eq:gammae}.

\item[2)] For given $X$, and $\Gamma_{\rm e}$, $Q$ can be estimated
  from Eq.\,\ref{eq:Qrecipe}.

\item[3)] For given $Q$, $W$ can be determined from Eq.\,\ref{eq:W},
  adopting reasonable values for $f_\rho$, and $D$. Note that the
  stability limit Eq.\,\ref{eq:goetz-limit} can be expressed in the
  form $Q/(f_\rho/D) < M/R$ (in solar units).

\item[4)] Finally, for given $W$, the radial extension $\Delta R$ can
  be computed from Eq.\,\ref{eq:DeltaRc}, or Eq.\,\ref{eq:DeltaRe}.

\end{itemize}

E.g., for our He models in Table\,\ref{tab:modparHE} (with $X=0$),
masses are typically of the order of $15\,M_\odot$, and core radii are
$\sim 1\,R_{\rm sun}$, i.e., $M/R \sim 15$ in solar units. This means
that an unstable situation, with $W=1$, is reached for $Q/(f_\rho/D)
\sim 15$ (cf.\,Eq.\,\ref{eq:W}).  For $f_\rho=2$, and $D=1$, this
corresponds to $Q \sim 30$.  According to Eq.\,\ref{eq:Qrecipe} this
value is reached for $\Gamma_{\rm e} \sim 0.4$, in agreement with our
models with $D=1$ in Table\,\ref{tab:modparH}.  For a large clumping
factor $D=16$, a value of $W=1$ is already reached for $Q \sim 2$,
suggesting an instability at $\Gamma_{\rm e} \sim 0.3$, in agreement
with our models with $D=16$ in Table\,\ref{tab:modparH}.

\begin{figure}[t!]
\parbox[b]{0.49\textwidth}{\includegraphics[scale=0.45]{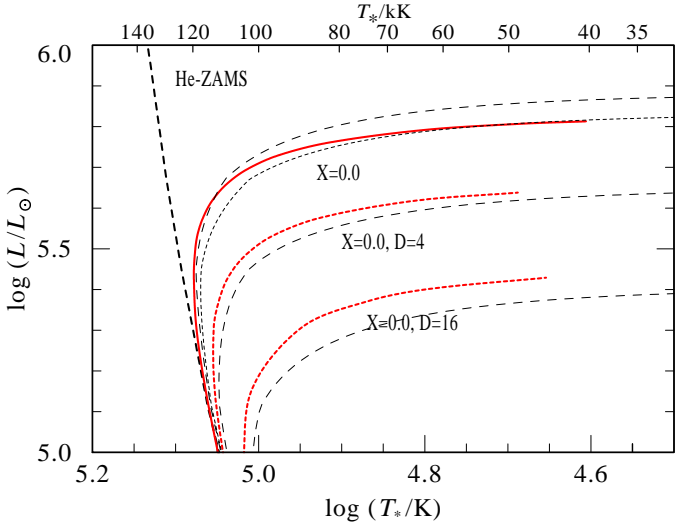}}  
\caption{HR diagram for pure He models (red), compared with our
  analytical relations. Long-dashed lines: analytical relation with
  $f_\rho = 2$, short-dashed line: analytical relation with $f_\rho =
  1.5$.}
\label{fig:HRD_anal}
\end{figure}

In Fig.\,\ref{fig:HRD_anal} we compare our analytical predictions,
based on core radii $R_{\rm c}$ from \citet{lan1:89}, with our
numerical models. Long-dashed lines indicate predicted HRD positions
within our analytical formalism for clumping factors $D=1$, 4, and 16,
and an adopted $f_\rho =2$.  The short-dashed line indicates the
results for $f_\rho =1.5$, and $D=1$.  Note that the use of the
He-ZAMS models by \citeauthor{lan1:89} introduces inconsistencies.
Nevertheless, our formalism describes the envelope inflation very
well, in particular its dependence on the clumping factor $D$.  Also
the uncertainty due to the adopted $f_\rho$ turns out to have only
moderate influence on the results.

\subsection{The mass of the surrounding shell $\Delta M$}
\label{sec:mcalc}

The mass $\Delta M$ of the dense shell surrounding the inflated zone
(cf.\,Fig.\,\ref{fig:density}) may be of interest, e.g.\ in relation
to the energy budget, and the timescales of variations in $\Delta R$.

To compute $\Delta M$ we use the following assumptions. 1) The radial
extension of the shell is small compared to the total stellar radius,
i.e., $r=R_{\rm e}$. 2) The mass of the shell is small compared to the
stellar mass, i.e., $m=M$. 3) The pressure $P$ at the stellar surface
is small compared to the pressure at the outer boundary of the
inflated region $P_{\rm e}$.

Under these assumptions, the equation of hydrostatic equilibrium
(Eq.\,\ref{eq:hydeq}) can be integrated directly, and we obtain
\begin{equation}
\Delta M = P_{\rm e}\frac{4\pi R_{\rm e}^4}{G M}.
\end{equation}
The mass of the shell thus simply follows from the fact that the
pressure force $P_{\rm e} 4\pi R_{\rm e}^2$ at the rim of the low
density region needs to be in equilibrium with the gravitational force
$GM\Delta M/R_{\rm e}^2$ on the shell. Again, $P_{\rm e}$ depends on
material properties. Using $P_{\rm e}\approx 0.4...0.9$$\cdot$$10^{6}\,{\rm
  dyn}/{\rm cm}^2$ from our models (cf.\,Table\,\ref{tab:modparHE} and
\ref{tab:modparH}), we obtain
\begin{equation}
  \label{eq:deltam}
  \Delta M/M_\odot \approx 0.5...1.0$$\cdot$$10^{-9}\, \frac{(R_{\rm e}/R_\odot)^4}{M/M_\odot}.
\end{equation}
$\Delta M$ thus mainly depends on the radius of the star. This is in
agreement with our model computations (Table\,\ref{tab:modparHE} and
\ref{tab:modparH}) where
we find $\Delta M$ as low as $10^{-9}\,M_\odot$ for compact WR stars,
whereas our larger, LBV-type models reach $\Delta M$ up to
$10^{-2}\,M_\odot$.

\subsection{Dependence on model assumptions}
\label{sec:assumptions}

{\changed The occurrence of an envelope inflation, involving a density
  inversion, has been reported in many previous works. Examples are
  the Wolf-Rayet models by \citet{sch1:96}, and models for very
  massive main sequence stars, as well as hydrogen-deficient carbon
  stars, by \citet{sai1:98}. What all these stars have in common, is
  their proximity to the Eddington limit. Especially the latter work,
  which discusses the connection of inflated envelopes to the
  strange-mode instability, demonstrates the universality of the
  principles discussed here, for stars covering the range of
  0.9--110\,$M_\odot$.

  On the other hand, the inflation effect has not been detected in
  many standard works on stellar evolution. E.g., in the rotating
  evolutionary models by \citet{mey1:03}, the $120\,M_\odot$ track
  enters the H-free WR stage with a mass of $23.2\,M_\odot$, and an
  effective core temperature of $T_\star = 123$\,kK. According to our
  results in Table\,\ref{tab:modparHE}, we would expect a considerable
  envelope inflation, with $T_\star \sim 60\,$kK, for the same object.
  In this section we discuss possible sources of such discrepancies,
  such as the influence of the imposed outer boundary condition
  (Sect.\,\ref{sec:OB}), dynamical terms on a secular timescale
  (Sect.\,\ref{sec:DYN}), dynamical terms introduced by mass-loss
  (Sect.\,\ref{sec:MDOT}), and the treatment of convection and
  turbulence (Sect.\,\ref{sec:conveff}).

}

\subsubsection{The effect of the outer boundary condition}
\label{sec:OB}

\begin{figure}[]
\parbox[b]{0.49\textwidth}{\includegraphics[scale=0.45]{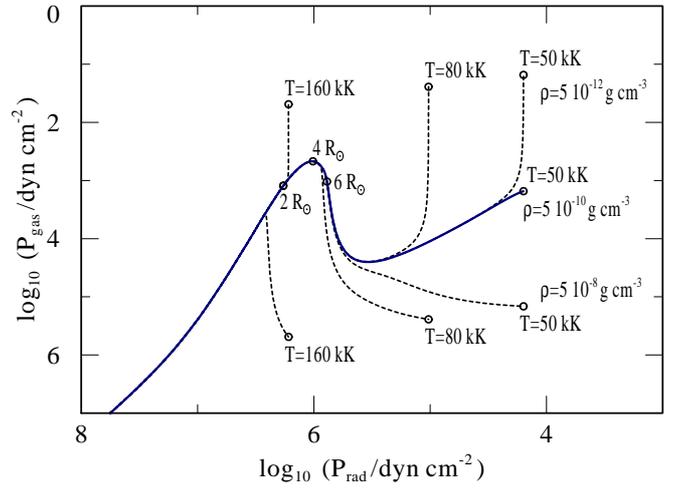}}  
\caption{\changed The effect of $\rho$ and $T$ at the outer boundary,
  on the envelope solution. The solid blue line indicates our standard
  $23\,M_\odot$ He-model in the $P_{\rm rad}$--$P_{\rm gas}$ plane.
  The dashed lines indicate solutions for which the boundary values
  for the density $\rho$, and the temperature $T$ have been changed.
  The inflated zone between 2 and 6\,$R_\odot$ (as indicated by the
  radius marks) is mainly affected by the choice of the temperature
  $T$ at the outer boundary. Temperatures $T \gtrsim 80$\,kK, can lead
  to a cut-off of the inflated zone.}
\label{fig:OB}
\end{figure}

{\changed In Fig.\,\ref{fig:OB} we illustrate the influence of the
  adopted boundary values for $\rho$, and $T$, on the envelope
  solution.  The blue line indicates our standard $23\,M_\odot$
  He-model, with boundary values $T=50$\,kK, and $\rho=5$$\cdot$$10^{-10}{\rm
    g}\,{\rm cm}^{-2}$. In addition we show six test models (dashed
  lines), for which we have changed $\rho$ to values of
  $5$$\cdot$$10^{-8}{\rm g}\,{\rm cm}^{-2}$ and $5$$\cdot$$10^{-12}{\rm g}\,{\rm
    cm}^{-2}$, for temperatures of $T=50$, 80, and 160\,kK.

  Firstly, the envelope solution turns out to be very robust with
  respect to changes of the outer boundary. For increasing $P_{\rm
    rad}$, i.e., towards the interior of the star, all test models
  merge into the standard solution.  However, boundary temperatures in
  the region of the Fe-opacity peak, i.e., temperatures that would lie
  {\em within} the inflated zone, lead to a cut-off of the inflated
  envelope. In Fig.\,\ref{fig:OB}, the inflated zone is indicated by
  radius marks (2--$6\,R_\odot$). Models with outer boundary
  temperatures $T \gtrsim 80$\,kK display envelopes where the inflated
  zone ends close to the respective outer boundary temperature, i.e.\
  the radii are significantly reduced.

  Boundary conditions that take the back-warming effect of an
  optically thick wind into account, as the ones discussed by
  \citet{sch1:96}, may thus inhibit an envelope inflation
  (\citeauthor{sch1:96} indeed detected a very similar effect in his
  model computations). In the same way, strong stellar winds with high
  temperatures at the wind base, as the ones described by
  \citet{gra1:05} for early WC subtypes, will prevent the formation of
  inflated envelopes. This is not surprising, as such winds are
  initiated by the hot Fe-opacity peak, i.e., the Fe-opacity peak lies
  within the wind. We thus expect a strong influence of the detailed
  wind physics on the inflation effect.

}

\subsubsection{Dynamical terms on a secular timescale}
\label{sec:DYN}

{\changed

The contraction/expansion of a star during its evolution can lead to
energy sources/sinks, in addition to the nuclear energy production
rate $\epsilon$, which is included in our models
(Eq.\,\ref{eq:econs}).  To estimate the importance of this effect, we
compare the gravitational energy $E_{\rm g}$, which is gained, or lost
on a timescale $\tau$, with the stellar luminosity $L$, and estimate in
this way a limiting timescale
\begin{equation}
  \tau =  \frac{\Delta E_{\rm g}}{L} = \frac{\Delta M(M-\Delta M)G}{L R}.
\label{eq:GRAV}
\end{equation}
Here $\Delta M$ is the mass of the layers involved in the
contraction/expansion, and $R$ is a typical radius.

Because Wolf-Rayet stars are very compact, and evolve on relatively
short timescales, such gravitational terms may play a role in their
evolution. However, as the masses of the inflated envelopes discussed
here are very small, these terms play no role for the inflation
effect.  For our 23.7\,$M_\odot$ He-model, which represents the most
extreme case of an inflated envelope, we estimate $\tau = 0.5$\,d,
based on the parameters from Table\,\ref{tab:modparHE}, and $R=R_{\rm
  c}$. We thus expect no influence of secular changes on the inflation
effect. In Sect.\,\ref{sec:lbv} we will discuss the case of LBVs, for
which gravitational terms indeed seem to play a role on timescales of
the observed S\,Dor-type variability.

For the {\em complete} radiative envelope of a typical WR star (here
we use the $15\,M_\odot$ model from Table\,\ref{tab:modparHE}, with
$\Delta M=1.5\,M_\odot$, and $R=0.4\,R_\odot$) we estimate $\tau =
5000$\,yr. Particularly in the end phases of the WR evolution,
gravitational terms may thus play an important role in determining the
$L/M$ ratio of WR stars, which is an important input for our
formalism.

}

\subsubsection{Dynamical terms due to mass-loss}
\label{sec:MDOT}

{\changed Because of the low densities, mass loss may introduce
  considerable velocity fields within the inflated envelopes, via the
  equation of continuity $\dot{M}=4\pi \rho \varv r^2$. The effect of
  the resulting dynamical terms on the envelope structure has been
  investigated by \citet{pet1:06}.}  They found that strong mass loss
can inhibit the formation of inflated envelopes, if the velocities
within the inflated zone exceed the local escape speed, i.e., when
$\varv({\rm d} \varv/{\rm d} r) \approx \varv^2/r$ has the same order
of magnitude as $GM/r^2$. This condition imposes an upper limit for
the mass-loss rate
\begin{equation}
\label{eq:mdotmax}
\dot{M}_{\rm 0} = 4\,\pi \rho_{\rm min} R_{\rm m}^2 \sqrt{\frac{MG}{R_{\rm m}}},
\end{equation}
where we have taken into account that the maximum velocity is reached
close to $R_{\rm m}=(R_{\rm e}+R_{\rm c})/2$, where
$\rho\approx\rho_{\rm min}$. An inspection of the dynamical term that
would result from the density structure within our own models,
suggests however that this limit should be lowered by $\sim$ 0.4\,dex.

{\changed In Table\,\ref{tab:mdotmax} we have compiled conservative
  estimates of $\dot{M}_{\rm 0}$, based on $\rho_{\rm min} =
  5$$\cdot$$10^{-11}\,{\rm g}/{\rm cm}^3$, for a set of WR, and LBV models.
  For the WR models with the lowest temperatures, which may be subject
  to an envelope inflation, we find $\log(\dot{M}_{\rm 0}/M_\odot {\rm
    yr}^{-1})= -4.2$...$-3.7$.  For comparison, spectroscopically
  determined mass-loss rates by \citet{ham1:06} reach values of up to
  $\log(\dot{M}/M_\odot {\rm yr}^{-1})= -4.3$...$-4.1$, dependent on
  $L$. In view of the involved uncertainties, it is thus not clear
  whether the envelope inflation of WR\,stars is affected by dynamical
  effects.}

\begin{table}[tp] \caption{Estimated limiting mass-loss rates $\dot{M}_{\rm 0}$
    according to Eq.\,\ref{eq:mdotmax}, based on an adopted value of
    $\rho_{\rm min} = 5$$\cdot$$10^{-11}\,{\rm g}/{\rm cm}^3$.
    \label{tab:mdotmax}}
  \centering
  \begin{tabular}{llllllll} \hline \hline
    \rule{0cm}{2.2ex}$\log(L)$ & $X$ & $T_\star$ & $M$ & $R_{\rm c}$& $R_\star$ & $R_{\rm m}$ &  $\log(\dot{M}_{\rm 0})$ \\
    $[L_\odot]$ & & $[{\rm kK}]$ & $[M_\odot$] & $[R_\odot]$& $[R_\odot]$ & $[R_\odot]$ &  $[M_\odot/{\rm yr}]$ \\
    \hline
    \rule{0cm}{2.2ex}5.8 & 0.0 & 100 & 24 & 1.7 & 2.6 & 2.2 & -4.3 \\
    5.8 & 0.0 &  75 & 24 & 1.7 &  4.7 & 3.2 & -4.1 \\
    5.8 & 0.0 &  50 & 24 & 1.7 & 10.6 & 6.2 & -3.7 \\
    5.3 & 0.0 & 100 & 12 & 1.0 &  1.5 & 1.3 & -4.8 \\
    5.3 & 0.0 &  75 & 12 & 1.0 &  2.6 & 1.8 & -4.6 \\
    5.3 & 0.0 &  50 & 12 & 1.0 &  6.0 & 3.5 & -4.2 \\
    \hline
    \rule{0cm}{2.2ex}6.15 & 0.36 &  17 & 73 & 14 & 137 & 59 & -1.9 \\
    6.12 & 0.36 &  28 & 70 & 14 & 51 & 30 & -2.4 \\
    \hline
  \end{tabular}
\end{table}

\subsubsection{Convection and turbulence}
\label{sec:conveff} 

As already noted in Sects.\,\ref{sec:inflation}, and \ref{sec:anainf},
the inefficiency of convection is a mandatory condition for the
occurrence of an envelope inflation. {\changed As the convective
  efficiency may be affected by details of the numerical
  implementation, or effects that are not taken into account in the
  standard mixing-length approach, we give basic physical arguments,
  why the convective efficiency should be low for the cases discussed
  in this work.}

Generally, the strong increase in opacity near the iron bump will lead
to the onset of convection through the usual Schwarzschild stability
criterion.  But for the models here such convection should be quite
inefficient in the sense that it can carry only a small fraction of
the local stellar flux $F$.  To see this, note that a robust upper
limit to the flux carried by convection is set by the {\em
  free-streaming} of internal energy $U$ at the local sound speed
$\varv_{\rm s}$, $F_{\rm fs} = \varv_{\rm s} U$.  In general, $U=
U_{\rm gas} + U_{\rm rad}$, but in the luminous stars here, and
particularly in the region around the iron bump, the radiative energy
dominates\footnote{Such convective transport of radiative energy
  assumes radiative diffusion from the convective cell takes longer
  than the turnover time. If this breaks down, the convective flux
  upper limit defined here would be even lower.}.  The associated
maximum fraction of stellar flux $F$ that could be carried by
convection is thus
\begin{equation}
  \frac{F_{\rm fs}}{F} \approx \varv_{\rm s} \frac{U_{\rm rad}}{F} 
  = \frac{4 \varv_{\rm s}}{c} \, \left ( \frac{T}{T_{\rm c}} \right )^{4} 
  \approx 0.0034 \left ( \frac{10^{5}\,{\rm K}}{T_{\rm c}} \right )^{4} 
  \, ,
 \label{eq:fconv}
\end{equation}
where $T_{\rm c}$ is the effective temperature for the
(pre-inflated) core radius $R_{\rm c}$.  The last equality applies the
characteristic iron bump temperature $T \approx 1.5 \times 10^{5}$\,K,
with associated sound speed $\varv_{\rm s} \approx 50$\,km/s.
This shows directly that for core effective temperatures appropriate
for Wolf-Rayet stars, convection can carry only a small fraction of
the total stellar flux, viz.\ 0.3\% for $T_{\rm c} = 10^{5}$\,K, and
about 5\% for $T_{\rm c} = 50,000$~K.  This implies that, even when
convection sets in, the bulk of the stellar flux is still transported
by radiative diffusion, thus justifying this basic simplifying
assumption in our analysis of envelope inflation.

{\changed We note that, due to the dominance of radiation pressure,
  and radiation energy in the inflated zone, the contributions from
  internal gas pressure/energy are generally negligible with respect
  to $F_{\rm fs}$. The same holds for turbulent pressure/energy, as
  $P_{\rm turb} = (2/3) u_{\rm turb} = (1/3) \rho a^2 {\cal M}^2$ is
  of the same order of magnitude as gas pressure/energy
  \citep[cf.,][p.\ 105]{mae1:09}. Here $u_{\rm turb}$ denotes the
  turbulent energy density, $a$ the sonic speed, and ${\cal M}$ the
  Mach number, which is expected to be lower than one.

  For low convective efficiency, the structure of inflated envelopes
  is solely determined by Eq.\,\ref{eq:ONE}, which is (in first order
  approximation) equivalent to the condition that $\Gamma = 1$. The
  latter only follows from material properties, so that the addition
  of turbulent pressure has almost no effect on the resulting envelope
  structure.}

\section{Implications}
\label{sec:implications}

Let us next discuss the implications of the predicted envelope
inflation for stars close to the Eddington limit, i.e., for WR stars
and LBVs. Both types of objects are extremely luminous, and are
thought to reside close to the Eddington limit. In
Sect.\,\ref{sec:wr}, we focus on the still unexplained WR locations in
the HR diagram.  In Sect.\,\ref{sec:lbv}, we discuss the possible
connection of the radius inflation to the S Doradus radius changes
characteristic of LBVs.

\subsection{The radius problem for Wolf-Rayet stars}
\label{sec:wr}

\begin{figure*}[]
\parbox[b]{0.99\textwidth}{\includegraphics[scale=0.7]{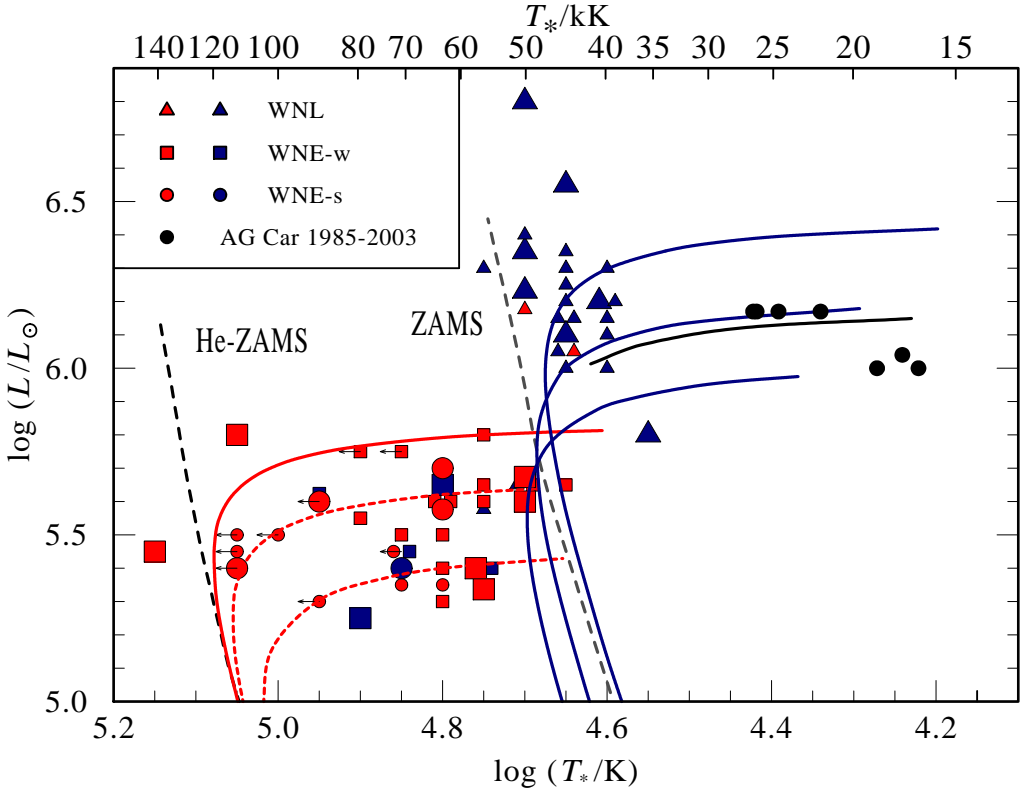}}  
\caption{Herzsprung-Russel diagram of the Galactic WN stars, and the
  LBV AG\,Car.  Red/blue symbols indicate observed HRD positions of WN
  stars from \citet{ham1:06} (blue: with hydrogen ($X>0.05$); red:
  hydrogen-free ($X<0.05$)). Black symbols indicate the HRD positions
  of AG\,Car throughout its S\,Dor Cycle from 1985--2003, according to
  \citet{gro1:09}.  Large symbols refer to stars with known distances
  from cluster/association membership. The symbol shapes indicate the
  spectral subtype (see inlet). Arrows indicate lower limits of
  $T_\star$ for stars with strong mass loss. The observations are
  compared to stellar structure models with/without hydrogen from this
  work (blue/red lines, cf.\,Fig.\,\ref{fig:hrdmod}), and for AG\,Car
  ($X=0.36$, black line).}
  \label{fig:hrdobs}
\end{figure*}

In Fig.\,\ref{fig:hrdobs}, we show an HR diagram for galactic WR
stars, as obtained from spectral analyses by \citet{ham1:06}. As
already highlighted by these authors there exists a radius problem for
H-free WR stars, namely that these objects are not located on the
theoretical He-main sequence. According to a population synthesis
performed by \citeauthor{ham1:06}, almost all WR stars with a hydrogen
surface abundance $X < 0.05$ are expected to display stellar
temperatures $T_\star$ in excess of 100\,kK. However, in the observed
HR diagram they cover a broad temperature range from 40--140\,kK, with
the majority below 100\,kK (red symbols in Fig.\,\ref{fig:hrdobs}).
The observed WR radii are thus larger than predicted by stellar
structure models \citep[e.g.,][]{lan1:89}.

At this point we note that the stellar temperatures $T_\star$, as
determined by \citet{ham1:06}, are not classical effective
temperatures (related to $\tau = 2/3$), but effective temperatures,
related to the (hydrostatic) surface radius $R_\star$, i.e.,
\begin{equation}
\label{eq:teff}
T_\star = \left( \frac{L}{4\pi\,\sigma_{\rm SB}\,R_\star^2} \right)^\frac{1}{4},
\end{equation}
with the Stefan-Boltzmann constant $\sigma_{\rm SB}$.  $R_\star$ is
the inner boundary radius of the employed atmosphere models, and is
typically located at a much higher optical depth of $\tau_{\rm
  max}\approx20$. As long as $R_\star$ is located in the hydrostatic
part of the wind, its value does not depend on the detailed choice of
$\tau_{\rm max}$ \footnote{The cases where $R_\star$ is located above
  the hydrostatic surface, are indicated as upper limits in
  Fig.\,\ref{fig:hrdobs}.}. The $T_\star$ determined by
\citeauthor{ham1:06} thus reflect actual surface radii, which are not
affected by the optical depth of the strong stellar winds of WR stars.
The low $T_\star$ in Fig.\,\ref{fig:hrdobs} thus display a true
inconsistency with respect to the stellar structure models.

This is particularly surprising, as H-free WR stars are very simple
objects. Due to their large convective cores, and high mass-loss rates
they are expected to be close to chemical homogeneity \citep{lan2:89}.
Even if the material in their He-burning cores is partly processed to
carbon and oxygen, the mean molecular weight stays almost constant
throughout the star. Moreover, the strong temperature sensitivity of
He-burning guarantees an almost fixed temperature in their centers.
The main unknown for the determination of the stellar structure is
thus the opacity. In fact, with the advent of the OPAL opacity data
\citep{igl1:96} and their strong Fe-opacity peaks, \citet{ish1:99}
were the first to predict the large envelope inflation that we also
describe in this work.

\citet{ish1:99} could however not explain the HRD positions of these
stars. In the galaxy, most H-free WR stars are located at luminosities
below $\log(L/L_\odot)\approx 5.8$, reaching down to
$\log(L/L_\odot)\approx 5.3$ (cf.\ Fig.\,\ref{fig:hrdobs}).
\citeauthor{ish1:99} demonstrated that unrealistically high
metallicities, up to $Z=0.1$, would be needed to achieve a significant
radius inflation in this regime.  An important result of the present
work is that the same effect can be achieved by the assumption of an
inhomogeneous (clumped) structure within the inflated zone
(cf.\,Sect.\ref{sec:clumping}). For moderate clumping factors, up to
$D=16$, our models cover the observed parameter range (red (dashed)
lines in Figs.\,\ref{fig:hrdmod} and \ref{fig:hrdobs}).  Notably,
similar clumping factors are determined spectroscopically, for the
winds of WR stars \citep[e.g.,][]{ham1:98}.

The metallicity dependence, as predicted by \citet{ish1:99} and
\citet{pet1:06}, may however still play an important role in the
determination of the temperatures of WR stars, as their subtype
distribution seems to depend on $Z$ \citep[e.g.,][]{cro1:07}. Later
spectral subtypes are typically found at higher $Z$, which is in line
with a stronger radius inflation due to the stronger Fe-opacity peak
at higher $Z$. This effect may however also depend on the increased
mass loss at higher $Z$ \citep[see][]{cro1:06}. {\changed We further
  note that, as discussed in Sect.\,\ref{sec:MDOT}, the mass-loss
  limit by \citet{pet1:06} may inhibit an envelope inflation for WR
  stars with strong mass-loss.}

\subsection{Radius changes of S Doradus-type LBVs}
\label{sec:lbv}

{\changed A defining property of LBVs are S Doradus-type radius
  variations on timescales of months to years. The inflation effect
  provides a reasonable explanation for radius variations on such
  timescales, as only small portions of the stellar envelope are
  affected. The underlying reason for the time dependence may be
  manifold. E.g., \citet{geo1:11} recently proposed that the
  variability of HD\,5980, an LBV-like star that temporarily shows a
  WR-type spectrum, may be triggered by a transition from a hot, to a
  cool wind base. This picture fits very nicely into the framework of
  the inflation effect. As discussed in Sect.\,\ref{sec:OB}, we expect
  that a hot, optically thick wind would inhibit a radial inflation,
  i.e., a wind transition as proposed by \citeauthor{geo1:11} could
  indeed lead to substantial radius changes.}

{\changed In the following we discuss the case of} AG\,Car, one of the
best-studied galactic S Doradus LBVs
\citep[e.g.,][]{lei1:94,sta1:01,vin1:02}.  More recently, AG Car has
been studied in detail throughout its full S\,Dor cycle (1985--2003),
by \citet{gro1:09,gro1:11}.  During this period the star increased its
radius from $58.5\,R_\odot$ to $120.4\,R_\odot$, corresponding to a
change in the stellar temperature from $T_\star = 26,450$\,K to
16,650\,K. Such radius and temperature changes are characteristic
aspects of the S\,Dor cycle, but they remain as yet unexplained by
stellar structure calculations \citep[see however][]{sto1:93}.  For
AG\,Car the luminosity was found to decrease from
$\log(L/L_\odot)=6.17$ to 6.0 between the S\,Dor visual minimum and
maximum phase.  Furthermore, \citet{gro1:09} determined a considerable
hydrogen deficiency for AG\,Car, with a hydrogen surface mass fraction
of $X=0.36$.

We have computed a series of homogeneous stellar structure models for
the relevant parameter range, with a hydrogen mass fraction of
$X=0.36$, and masses ranging from 60--$73\,M_\odot$
(cf.\,Table\,\ref{tab:modparH}). These models display a similar
luminosity range as AG\,Car, and partly show very low $T_\star$, down
to $\sim$\,17,000\,K, due to a substantial radius inflation. A
comparison with the observed HRD positions of AG\,Car is shown in
Fig.\,\ref{fig:hrdobs}. Notably, during the minimum phases of the
S\,Dor cycle, AG\,Car (indicated by black symbols) shows a very good
agreement with our homogeneous models (black dashed line), i.e., its
position in the HR diagram can be explained by a chemically
homogeneous star with $M \sim 70\,M_\odot$, and a substantial radius
inflation.

The core of such a star would be relatively small, with $R_{\rm c}\sim
14\,R_\odot$, corresponding to an effective core temperature $T_{\rm
  c}\sim 53$\,kK (cf.\,Table\,\ref{tab:modparH}).  Due to the envelope
inflation, the star would display a core-halo structure, as described
in Sect.\,\ref{sec:inflation}. According to Eq.\,\ref{eq:deltam},
the mass $\Delta M$, of the outer shell around the inflated envelope,
is expected to change throughout the S\,Dor cycle. For the smallest
radius ($58.5\,R_\odot$), we obtain $\Delta M \sim 10^{-4} M_\odot$,
while $\Delta M \sim 2$$\cdot$$10^{-3} M_\odot$ for the largest extension
($120.4\,R_\odot$).  The energy to lift this material from the core
radius to the stellar surface is $\Delta E_{\rm g} \sim \Delta M M
G/{R_{\rm c}} = 3.8$$\cdot$$10^{46}\,{\rm erg}$. This value compares well
with the 'missing energy' due to the decrease in luminosity $\Delta
E_{\rm lum} = 1.83$$\cdot$$10^{39} {\rm erg/s} \times 575 {\rm d} =
9.1$$\cdot$$10^{46} {\rm erg}$ \citep[cf.\ ][]{gro1:09}. {\changed We thus
  find that within the inflated envelope, the S\,Dor-type variability
  introduces dynamical effects as the ones discussed in
  Sect.\,\ref{sec:DYN}.}

The good agreement with our models suggests that the extension of the
envelope of AG\,Car {\changed may be explained by the inflation
  effect.} The mass-loss rate of AG\,Car \citep[$\log(\dot{M}/M_\odot
{\rm yr}^{-1}) < -4.2$,][]{gro1:09} lies well below the upper
mass-loss limit discussed in Sect.\ref{sec:MDOT} (cf.\
Table\,\ref{tab:mdotmax}), implying a stable, static configuration
that is not affected by {\changed dynamical effects due to mass loss.}
Based on our chemically homogeneous models, AG\,Car lies close to the
Eddington limit, with an Eddington factor of $\Gamma_{\rm e} = 0.39$.
According to Eq.\,\ref{eq:Qrecipe}, this results in a $Q$ parameter
of $\sim 9$, and with Eq.\,\ref{eq:W} we obtain $W_{\rm c} \sim
(9/f_\rho) (R_{\rm c}/R_\odot)/(M/M_\odot) \sim 0.8$ (where we have
adopted $f_\rho=2.2$ from\,Table\,\ref{tab:modparH}). The star is thus
very close to the instability limit ($W_{\rm c}=1$), as discussed in
Sect.\,\ref{sec:rcalc}.

One possible mechanism that could lead to an unstable situation 
involving $W_{\rm c}>1$, is the reduction of the effective stellar mass through
rotation. \citet{gro1:11} found that the observed rotational velocity
of AG\,Car declines from $\varv_{\rm rot} \sin i = 220\,$km/s in the
minimum phase, to 90\,km/s for the largest extension. For these values
the ratio of centrifugal to gravitational acceleration at the stellar
surface decreases from $\varv_{\rm rot}^2 R_\star/(GM) = 0.21/\sin^2
i$, to $0.07/\sin^2 i$. The corresponding reduction of the effective
stellar mass by 20\% in the minimum phase, could thus just lead to a
situation where $W_{\rm c} \approx 1$. In this case the outer stellar
envelope would become unstable, and could start expanding. 
The reduction of the rotational velocity during the expansion would 
lead to $W_{\rm c} <1$, and a re-stabilization of the envelope.

\citet{gro2:09,gro1:11} proposed a somewhat similar scenario for
AG\,Car, where the Eddington limit is exceeded due to rotation. In
this work, we have demonstrated that already for moderate Eddington
factors of $\Gamma_{\rm e} \sim 0.4$\footnote{Note that, by
  definition, $\Gamma_{\rm e}$ only takes the opacity due to free
  electrons into account.}  AG\,Car may form an inflated stellar
envelope that is dominated by radiation. Within this envelope the star
adjusts very precisely to the Eddington limit ($\Gamma = 1$). In our
scenario, the instability that potentially induces the S\,Dor cycle is
due to the fact that the specific internal energy due to radiation
exceeds the gravitational potential, with the stellar envelope thus
becoming unbound.

{\changedA We want to point out that the formation of inflated
  envelopes does not depend on the assumption of chemical homogeneity
  in our numerical models. The instability follows from our formalism
  in Sect.\,\ref{sec:anal}, and is thus independent of the internal
  structure of the star. For AG\,Car, the observed luminosity
  ($\log(L/L_\odot)=6.17$), and surface hydrogen mass fraction
  ($X=0.36$) imply a mass of $74\,M_\odot$ for the case of chemical
  homogeneity \citep[cf.][]{gra1:11}. In reality, AG\,Car is likely
  more chemically enriched in the core, i.e., its mass is lower, and
  its core radius larger. This would lead to an even earlier onset of
  the inflation effect, and its instability. As the evolutionary
  status of LBVs is not well known, it is however not possible to
  obtain reliable mass estimates for this type of object.}

\section{Conclusions}
\label{sec:conclusions}

In the present work, we have discussed the possible formation of
radially inflated stellar envelopes for stars approaching the
Eddington limit, as originally proposed by \citet{ish1:99}, and
\citet{pet1:06}. {\changedA In addition to numerical models, we could
  provide an analytic description of this process, leading to the
  discovery of a new instability limit, and a clumping dependence.
  Both effects turn out to be of profound importance. While the former
  may be connected to S Doradus-type instabilities in LBVs, the latter
  can account for the large observed radii of Galactic WR stars, and
  thus resolve the long standing WR radius problem.}

{\changedA Within our analytical approach we could describe the
  inflation effect in terms of a dimensionless parameter $W$}
(Eqs.\,\ref{eq:Qrecipe}, \ref{eq:W}) that characterizes the
ratio between internal and gravitational energy at the base of the
envelope. $W$ can be estimated on the basis of opacity tables and a
combination of stellar parameters. The envelope inflation occurs when
$W$ approaches one. For $W\ge 1$, we find that the envelope becomes
gravitationally unbound, i.e., there exists no static solution.  A
prerequisite for envelope inflation is the proximity of the star to
the Eddington limit, with Eddington factors of $\Gamma_{\rm
  e}>0.3$--0.35, at solar metallicity.  For a given chemical
composition and $\Gamma_{\rm e}$, the resulting stellar radius depends
on the ratio $M/R$, and on the degree of (in)homogeneity of the
material within the envelope (characterized by a clumping factor $D$).

Envelope inflation has a strong impact on the effective temperatures
of stars in the upper HR diagram. For Wolf-Rayet stars it can account
for the 'radius problem', i.e., that the observed stellar temperatures
of H-free WR stars are much lower than predicted by canonical
stellar structure models. To reproduce the observed HRD positions of
these objects, it is necessary to assume that the material within the
envelope is clumped. The resulting clumping factors lie in the range
of $D=1$--16, in notable agreement with typical clumping factors
detected in the winds of WR stars \citep[e.g.,][]{ham1:98}.  As the
envelope inflation is expected to depend on metallicity
\citep{ish1:99,pet1:06}, it may also account for the observed
$Z$-dependence of the spectral subtype distribution of WR stars, i.e.,
that WR stars in high-$Z$ environments show lower effective
temperatures \citep[e.g.,][]{cro1:07}.

For the luminous blue variable AG\,Car we suggest that the observed
HRD position of a $\sim$\,$70\,M_\odot$ star, with a relatively
compact core, might be subject to substantial envelope inflation.
During its S\,Dor cycle, AG\,Car is found to increase 
its radius by a factor of two, while its luminosity decreases by a 
factor of 1.5. The luminosity
decrease might be explained as a result of the formation of a dense
outer shell with a mass of $\sim 2$$\cdot$$10^{-3}\,M_\odot$ as predicted by
our models. The energy needed to lift this material from the core to
the outer radius is of the same order of magnitude as the 'missing
energy' due to the decrease in luminosity. Similar to the mechanism
proposed by \citet{gro2:09,gro1:11}, the S\,Dor variability of AG\,Car
may be due to the effect of stellar rotation, which reduces the
effective stellar mass, and thus leads to an instability with $W>1$,
that forces the star to expand.

We conclude that the stellar effective temperatures in the upper HR
diagram are potentially strongly affected by the inflation effect.
The peculiar structure of the inflated envelopes, with an almost void,
radiatively dominated region, beneath a thin and dense shell could
mean that many, rather very compact stars, are hidden below inflated
envelopes, and thus display much lower effective temperatures to the
observer. Also compact, fast rotating stars may be hidden this way.
This may particularly affect massive stars just before the final
collapse, and the question of whether they form a SN\,Ibc, or a GRB
\citep[for the latter smaller core radii are inferred than the
observed WR radii, cf.][]{cui1:10}.  The complex envelope structure
may also affect the early X-ray afterglow of GRBs
\citep[e.g.][]{li1:07}.

\begin{acknowledgements}
  We thank A.\ Maeder for a very constructive referee report, and
  S.-C.\ Yoon for helpful discussions about evolutionary effects in
  WR\,stars.
\end{acknowledgements}


\begin{thebibliography}{46}
\expandafter\ifx\csname natexlab\endcsname\relax\def\natexlab#1{#1}\fi

\bibitem[{{Clark} {et~al.}(2009){Clark}, {Crowther}, {Larionov}, {Steele},
  {Ritchie}, \& {Arkharov}}]{cla1:09}
{Clark}, J.~S., {Crowther}, P.~A., {Larionov}, V.~M., {et~al.} 2009, \aap, 507,
  1555

\bibitem[{{Crowther}(2007)}]{cro1:07}
{Crowther}, P.~A. 2007, ARA\&A, 45, 177

\bibitem[{{Crowther} \& {Hadfield}(2006)}]{cro1:06}
{Crowther}, P.~A. \& {Hadfield}, L.~J. 2006, A\&A, 449, 711

\bibitem[{{Cui} {et~al.}(2010){Cui}, {Liang}, {Lv}, {Zhang}, \& {Xu}}]{cui1:10}
{Cui}, X.-H., {Liang}, E.-W., {Lv}, H.-J., {Zhang}, B.-B., \& {Xu}, R.-X. 2010,
  \mnras, 401, 1465

\bibitem[{{Eddington}(1918)}]{edd1:18}
{Eddington}, A.~S. 1918, \apj, 48, 205

\bibitem[{{Gal-Yam} \& {Leonard}(2009)}]{gal1:09}
{Gal-Yam}, A. \& {Leonard}, D.~C. 2009, \nat, 458, 865

\bibitem[{{Georgiev} {et~al.}(2011){Georgiev}, {Koenigsberger}, {Hillier},
  {Morrell}, {Barb{\'a}}, \& {Gamen}}]{geo1:11}
{Georgiev}, L., {Koenigsberger}, G., {Hillier}, D.~J., {et~al.} 2011, \aj, 142,
  191

\bibitem[{{Glatzel}(2008)}]{gla1:08}
{Glatzel}, W. 2008, in Astronomical Society of the Pacific Conference Series,
  Vol. 391, Hydrogen-deficient stars, ed. {A.~Werner \& T.~Rauch} (San
  Francisco: Astronomical Society of the Pacific), 307

\bibitem[{{Glatzel} \& {Kaltschmidt}(2002)}]{gla1:02}
{Glatzel}, W. \& {Kaltschmidt}, H.~O. 2002, MNRAS, 337, 743

\bibitem[{{Gr{\"a}fener} \& {Hamann}(2005)}]{gra1:05}
{Gr{\"a}fener}, G. \& {Hamann}, W.-R. 2005, A\&A, 432, 633

\bibitem[{{Gr{\"a}fener} \& {Hamann}(2008)}]{gra1:08}
{Gr{\"a}fener}, G. \& {Hamann}, W.-R. 2008, A\&A, 482, 945

\bibitem[{Gr\"afener {et~al.}(2002)Gr\"afener, Koesterke, \& Hamann}]{gra1:02}
Gr\"afener, G., Koesterke, L., \& Hamann, W.-R. 2002, A\&A, 387, 244

\bibitem[{{Gr{\"a}fener} {et~al.}(2011){Gr{\"a}fener}, {Vink}, {de Koter}, \&
  {Langer}}]{gra1:11}
{Gr{\"a}fener}, G., {Vink}, J.~S., {de Koter}, A., \& {Langer}, N. 2011, \aap,
  535, A56

\bibitem[{{Groh} {et~al.}(2009{\natexlab{a}}){Groh}, {Damineli}, {Hillier},
  {Barb{\'a}}, {Fern{\'a}ndez-Laj{\'u}s}, {Gamen}, {Mois{\'e}s}, {Solivella},
  \& {Teodoro}}]{gro2:09}
{Groh}, J.~H., {Damineli}, A., {Hillier}, D.~J., {et~al.} 2009{\natexlab{a}},
  \apjl, 705, L25

\bibitem[{{Groh} {et~al.}(2011){Groh}, {Hillier}, \& {Damineli}}]{gro1:11}
{Groh}, J.~H., {Hillier}, D.~J., \& {Damineli}, A. 2011, \apj, 736, 46

\bibitem[{{Groh} {et~al.}(2009{\natexlab{b}}){Groh}, {Hillier}, {Damineli},
  {Whitelock}, {Marang}, \& {Rossi}}]{gro1:09}
{Groh}, J.~H., {Hillier}, D.~J., {Damineli}, A., {et~al.} 2009{\natexlab{b}},
  \apj, 698, 1698

\bibitem[{{Hamann} {et~al.}(2006){Hamann}, {Gr{\"a}fener}, \&
  {Liermann}}]{ham1:06}
{Hamann}, W.-R., {Gr{\"a}fener}, G., \& {Liermann}, A. 2006, A\&A, 457, 1015

\bibitem[{{Hamann} \& {Koesterke}(1998)}]{ham1:98}
{Hamann}, W.-R. \& {Koesterke}, L. 1998, A\&A, 335, 1003

\bibitem[{{Hansen} \& {Kawaler}(1994)}]{han1:94}
{Hansen}, C.~J. \& {Kawaler}, S.~D. 1994, {Stellar Interiors. Physical
  Principles, Structure, and Evolution.} (Springer, Secaucus, New Jersey,
  U.S.A)

\bibitem[{{Heger} {et~al.}(2003){Heger}, {Fryer}, {Woosley}, {Langer}, \&
  {Hartmann}}]{heg1:03}
{Heger}, A., {Fryer}, C.~L., {Woosley}, S.~E., {Langer}, N., \& {Hartmann},
  D.~H. 2003, ApJ, 591, 288

\bibitem[{{Humphreys} \& {Davidson}(1994)}]{hum1:94}
{Humphreys}, R.~M. \& {Davidson}, K. 1994, \pasp, 106, 1025

\bibitem[{{Iglesias} \& {Rogers}(1996)}]{igl1:96}
{Iglesias}, C.~A. \& {Rogers}, F.~J. 1996, ApJ, 464, 943

\bibitem[{{Ishii} {et~al.}(1999){Ishii}, {Ueno}, \& {Kato}}]{ish1:99}
{Ishii}, M., {Ueno}, M., \& {Kato}, M. 1999, PASJ, 51, 417

\bibitem[{{Kotak} \& {Vink}(2006)}]{kot1:06}
{Kotak}, R. \& {Vink}, J.~S. 2006, \aap, 460, L5

\bibitem[{{Langer}(1989{\natexlab{a}})}]{lan2:89}
{Langer}, N. 1989{\natexlab{a}}, A\&A, 220, 135

\bibitem[{{Langer}(1989{\natexlab{b}})}]{lan1:89}
{Langer}, N. 1989{\natexlab{b}}, A\&A, 210, 93

\bibitem[{{Langer} {et~al.}(1994){Langer}, {Hamann}, {Lennon}, {Najarro},
  {Pauldrach}, \& {Puls}}]{lan1:94}
{Langer}, N., {Hamann}, W.-R., {Lennon}, M., {et~al.} 1994, A\&A, 290, 819

\bibitem[{{Leitherer} {et~al.}(1994){Leitherer}, {Allen}, {Altner}, {Damineli},
  {Drissen}, {Idiart}, {Lupie}, {Nota}, {Robert}, {Schmutz}, \&
  {Shore}}]{lei1:94}
{Leitherer}, C., {Allen}, R., {Altner}, B., {et~al.} 1994, \apj, 428, 292

\bibitem[{{Li}(2007)}]{li1:07}
{Li}, L.-X. 2007, \mnras, 375, 240

\bibitem[{{Maeder}(2009)}]{mae1:09}
{Maeder}, A. 2009, {Physics, Formation and Evolution of Rotating Stars}
  (Springer Berlin Heidelberg)

\bibitem[{{Meynet} \& {Maeder}(2003)}]{mey1:03}
{Meynet}, G. \& {Maeder}, A. 2003, A\&A, 404, 975

\bibitem[{{Modjaz} {et~al.}(2009){Modjaz}, {Li}, {Butler}, {Chornock},
  {Perley}, {Blondin}, {Bloom}, {Filippenko}, {Kirshner}, {Kocevski},
  {Poznanski}, {Hicken}, {Foley}, {Stringfellow}, {Berlind}, {Barrado y
  Navascues}, {Blake}, {Bouy}, {Brown}, {Challis}, {Chen}, {de Vries},
  {Dufour}, {Falco}, {Friedman}, {Ganeshalingam}, {Garnavich}, {Holden},
  {Illingworth}, {Lee}, {Liebert}, {Marion}, {Olivier}, {Prochaska},
  {Silverman}, {Smith}, {Starr}, {Steele}, {Stockton}, {Williams}, \&
  {Wood-Vasey}}]{mod1:09}
{Modjaz}, M., {Li}, W., {Butler}, N., {et~al.} 2009, \apj, 702, 226

\bibitem[{{Petrovic} {et~al.}(2006){Petrovic}, {Pols}, \& {Langer}}]{pet1:06}
{Petrovic}, J., {Pols}, O., \& {Langer}, N. 2006, \aap, 450, 219

\bibitem[{{Saio} {et~al.}(1998){Saio}, {Baker}, \& {Gautschy}}]{sai1:98}
{Saio}, H., {Baker}, N.~H., \& {Gautschy}, A. 1998, \mnras, 294, 622

\bibitem[{{Schaerer}(1996)}]{sch1:96}
{Schaerer}, D. 1996, A\&A, 309, 129

\bibitem[{{Schaller} {et~al.}(1992){Schaller}, {Schaerer}, {Meynet}, \&
  {Maeder}}]{sch1:92}
{Schaller}, G., {Schaerer}, D., {Meynet}, G., \& {Maeder}, A. 1992, A\&AS, 96,
  269

\bibitem[{{Smith} {et~al.}(2007){Smith}, {Li}, {Foley}, {Wheeler}, {Pooley},
  {Chornock}, {Filippenko}, {Silverman}, {Quimby}, {Bloom}, \&
  {Hansen}}]{smi2:07}
{Smith}, N., {Li}, W., {Foley}, R.~J., {et~al.} 2007, \apj, 666, 1116

\bibitem[{{Smith} {et~al.}(2004){Smith}, {Vink}, \& {de Koter}}]{smi1:04}
{Smith}, N., {Vink}, J.~S., \& {de Koter}, A. 2004, \apj, 615, 475

\bibitem[{{Stahl} {et~al.}(2001){Stahl}, {Jankovics}, {Kov{\'a}cs}, {Wolf},
  {Schmutz}, {Kaufer}, {Rivinius}, \& {Szeifert}}]{sta1:01}
{Stahl}, O., {Jankovics}, I., {Kov{\'a}cs}, J., {et~al.} 2001, \aap, 375, 54

\bibitem[{{Stothers} \& {Chin}(1993)}]{sto1:93}
{Stothers}, R.~B. \& {Chin}, C.-W. 1993, \apjl, 408, L85

\bibitem[{{Vink}(2009)}]{vin1:09}
{Vink}, J.~S. 2009, ArXiv e-prints, 0905.3338 (in press)

\bibitem[{{Vink} \& {de Koter}(2002)}]{vin1:02}
{Vink}, J.~S. \& {de Koter}, A. 2002, \aap, 393, 543

\bibitem[{{Vink} {et~al.}(2011){Vink}, {Muijres}, {Anthonisse}, {de Koter},
  {Gr{\"a}fener}, \& {Langer}}]{vin1:11}
{Vink}, J.~S., {Muijres}, L.~E., {Anthonisse}, B., {et~al.} 2011, \aap, 531,
  A132

\bibitem[{{Woosley} \& {Heger}(2006)}]{woo1:06}
{Woosley}, S. \& {Heger}, A. 2006, ApJ, 637, 914

\bibitem[{{Yoon} \& {Langer}(2005)}]{yoo1:05}
{Yoon}, S.-C. \& {Langer}, N. 2005, A\&A, 443, 643

\bibitem[{{Yungelson} {et~al.}(2008){Yungelson}, {van den Heuvel}, {Vink},
  {Portegies Zwart}, \& {de Koter}}]{yun1:08}
{Yungelson}, L.~R., {van den Heuvel}, E.~P.~J., {Vink}, J.~S., {Portegies
  Zwart}, S.~F., \& {de Koter}, A. 2008, \aap, 477, 223

\end{thebibliography}

\end{document}